\documentclass[12pt]{article}
\usepackage[utf8]{inputenc}
\usepackage[T1]{fontenc}
\usepackage{mathpazo}
\usepackage[margin=1in]{geometry}
\usepackage{amsmath, amssymb, amsthm}
\usepackage{setspace}
\usepackage{float}
\usepackage{placeins}
\usepackage{xcolor}
\usepackage[round,semicolon]{natbib}
\usepackage{xurl}
\usepackage{microtype}
\usepackage{enumitem}
\usepackage{graphicx}
\usepackage{subcaption}
\usepackage{booktabs}
\usepackage[colorlinks=true,
            linkcolor={blue!70!black},
            citecolor={blue!40!black},
            urlcolor={blue!40!black}]{hyperref}
\usepackage{tabularx}
\usepackage{array}

\newtheorem{result}{Result}

\theoremstyle{definition}
\newtheorem{definition}{Definition}
\newtheorem{assumption}{Assumption}
\theoremstyle{plain}

\newcolumntype{D}{>{\hsize=1.45\hsize\raggedright\arraybackslash}X}
\newcolumntype{S}{>{\hsize=0.55\hsize\raggedright\arraybackslash}X}

\graphicspath{{figures/}}

\title{\textbf{AI, Automation, and Fiscal Pressure}\thanks{We thank participants at workshops organized by AISCAF and WASP-HS for useful comments. The project is funded by the WASP-HS grant ``AI and the Future of Taxes''.}}
\author{
    Spencer Bastani\thanks{Institute for Evaluation of Labour Market and Education Policy (IFAU) and Department of Economics, Uppsala University; Research Institute of Industrial Economics (IFN). Email: \texttt{spencer.bastani@ifau.uu.se}.}
    \and
    Georg Graetz\thanks{Department of Economics, Uppsala University. Email: \texttt{georg.graetz@nek.uu.se}.}
    \and
    Bernat Pla\thanks{Department of Economics, Uppsala University. Email: \texttt{bernat.pla@nek.uu.se}.}
    \and
    Oskar Nordstr\"om Skans\thanks{Department of Economics, Uppsala University. Email: \texttt{oskar.nordstrom\_skans@nek.uu.se}.}
}
\date{November 24, 2026}

\begin{document}

\maketitle

\begin{abstract}
	\noindent

We study how automation affects public finances in countries that (i) tax labor at a higher rate than capital and (ii) fund services such as health care and education from tax revenues rather than user fees. We embed task-level automation within a two-sector general equilibrium model, which we calibrate to match data from EU countries. Automation shifts income toward more lightly taxed capital. If public services follow consumer demand, the budget deficit worsens at unchanged tax rates, even if automation is as fast in the public-service sector as in the private sector. In our EU-27 benchmark scenario, where automation proceeds faster in the market sector, the deficit rises by nearly one percentage point of value added as the economy grows by twelve percent due to automation. 


\smallskip\noindent \textbf{JEL-Codes}: E25, E62, H20, H42, O33

\end{abstract}

\clearpage
\section{Introduction}
In much of the developed world, public finances are under increasing pressure due to expanding welfare states and slow economic growth. Faster growth is widely seen as a natural relief: a larger economy can generate larger tax revenues to help finance existing commitments without raising tax rates. In the near future, advances in artificial intelligence (AI) are expected to be a key engine of economic growth as AI enables automation of a wide set of tasks currently performed by human labor \citep{brynjolfsson2021productivity, aghion2019artificial}. But growth arising from AI-driven automation is likely to have fiscal effects beyond the growth of economic activity, as it also shifts how income is distributed between factors and
how costly it is to produce public services. The goal of this paper is to provide a quantitative model of fiscal pressure under AI-driven automation. The model is built to be relevant for the European setting where most countries satisfy two key characteristics. First, labor is taxed at a higher rate than capital. Second, services such as health care and education are funded (though not necessarily run) by the state rather than user fees.

Our aim is to assess whether AI-driven automation will raise or lower the government's primary budget deficit---in other words, whether AI heightens or eases fiscal pressure---and to quantify how large these effects are under different scenarios.  The key issues at play emerge from simple budgetary arithmetic. The primary deficit ($b$) as a share of value added  satisfies
\begin{equation}\label{eq:budget-deficit}
    b = \underbrace{s_E}_{\substack{\text{public-service} \\ \text{expenditure}}} 
    - \underbrace{\tau_L s_L}_{\substack{\text{labor} \\ \text{tax revenue}}} 
    - \underbrace{\tau_K(1-s_L)}_{\substack{\text{capital} \\ \text{tax revenue}}} 
    - \underbrace{\tau_c(1-s_E)}_{\substack{\text{consumption} \\ \text{tax}}} 
    - \underbrace{\tilde{\tau}}_{\text{residual}},
\end{equation}
where all right-hand side terms are also expressed in percent of value added.\footnote{The primary budget deficit equals the total deficit minus interest payments on public debt.} Here, $s_L$ is the labor share of value added, the consumption tax is only levied on private consumption, and $\tilde{\tau}$ captures all other taxes net of transfers and other government spending. Holding tax rates and the residual term $\tilde{\tau}$ fixed, we focus on two channels. The first is the tax-base channel. When automation lowers the labor share, the deficit worsens because labor is taxed at a higher rate than capital. The second is the public-service expenditure channel. To offset the tax-base effect, the public-service share must decline. Thus, to assess whether AI may ease fiscal pressure, we require a theory which helps us understand and quantify the determinants of that share.

Our general equilibrium model embeds the standard task model of production in an economy with a tax-funded public-service sector and a perfectly competitive market sector. Capital can only perform a subset of tasks, but technological advances such as AI expand the range of tasks that can be automated. In our baseline model, where capital supply is perfectly elastic, automation raises output and wages, while labor's share of income falls. Public services experience an increase in their production costs relative to the market sector for two reasons: first, due to the rise in wages, as they are more labor intensive; and second, because automation likely proceeds more slowly, due to task mix, a preference for human delivery, or regulatory barriers.\footnote{For instance, using the AI exposure measure of \citet{engberg2024ai}, \citet{konjunkturradet2026} find that AI has less potential to raise efficiency in the Swedish welfare sector than in the rest of the economy, although the gap narrowed between 2013 and 2022.} In short, public services are hit disproportionately by higher wages while being at a disadvantage in reaping the benefits of AI-driven automation. 

A key question is how public-service provision responds as the economy grows and public services become relatively more costly. Our analysis explores two polar cases. As one polar case, public services are held fixed in real terms, generating a decline in the expenditure share ($s_E$) that counteracts the erosion of the tax base. This case shows whether AI-driven growth can pay for existing public-service commitments at unchanged tax rates, but it does not accommodate growing demand for services as incomes rise. As our main case of interest, we therefore instead assume that public-service provision fully responds to changes in relative production costs, accommodating the entire change in household demand for services. Under the reasonable assumption that the elasticity of substitution between public services and market consumption is below one, the rise in the relative cost of services will raise the expenditure share ($s_E$)---the well-known Baumol effect. In this case, the public-service share does not fall, so the tax-base erosion is not offset, and a rising expenditure share adds further fiscal pressure.

We quantify these effects by calibrating our model to the EU-27 as a whole and to each member state separately using national accounts and tax data from 2023. We consider a range of automation scenarios in which each automated task initially saves 20 percent of its cost, an assumption motivated by the task-level cost savings discussed by \citet{acemoglu2025simple}. In our benchmark scenario, the task share performed by capital increases by twice as much in the market sector as in the public-service sector---5 and 2.5 percentage points (pp) respectively---with alternative ratios reported in a scenario grid.

For the EU-27, the deficit increases by nearly 1pp of value added under \textit{demand-responsive provision} in our benchmark scenario, even as output grows by 12 percent. Almost all of the increase comes from the tax base. The Baumol effect adds less than 0.1pp. The key mechanism is instead that the public-service share \textit{does not fall} while the tax base is eroded. The deficit increases in every demand-responsive scenario we consider, even when automation is as fast in the public-service sector as in the market sector. Qualitatively, the results are similar across all 27 EU countries, although the magnitudes vary with the size of the labor--capital tax differential. The increases are more than three times as large in Slovakia and Greece as in Luxembourg and the Netherlands. Under \textit{fixed provision}, in contrast, the deficit falls by 1.5pp of value added, because the public-service share declines by about 2pp, which more than offsets the tax-base effect. This consolidation arises in a scenario where output grows by 14 percent while the real public-service bundle is held constant by assumption.

The main message is robust across a wide range of sensitivity checks. These sensitivity checks do, however, provide important insights into the underlying forces. The benchmark results are largely unchanged if we assume that capital income in the public-service sector is not taxed (with the capital-tax rate raised to preserve initial capital-tax revenue). If, on the other hand, we assume a fixed capital stock instead of perfectly elastic capital supply, signs remain unaltered in the benchmark scenario but effects become more muted, and wages fall instead of rising. 

With foreign ownership of capital, and in particular if the adjustment in the capital stock required to keep the rental rate fixed is entirely due to foreign investment, fiscal pressure rises even under fixed provision (using the same scenario as discussed above). In this case, the public expenditure share falls just as in the baseline model, but the fall no longer offsets the loss of revenue, as the income paid abroad escapes the consumption tax and, under residence-based taxation, the capital tax as well. How \textit{much} fiscal pressure rises depends on where capital income is taxed.

The fiscal-pressure measure defined in equation~\eqref{eq:budget-deficit} places cash transfers in the residual term. By holding this term fixed, we implicitly assume that transfers grow at the same rate as GDP. Our results understate fiscal pressure if transfers instead grow faster than GDP, for instance due to an increase in benefit payments to help workers cope with labor market disruption, which we do not model. In contrast, transfers may be subject to more frugal indexation rules, which we analyze in an extension. If benefits are fixed in real terms or indexed to wages, their share of value added falls by more than fiscal pressure rises under demand-responsive provision in every EU-27 scenario. Falling transfer shares can therefore offset the increase in fiscal pressure. This offset arises because these rules tie benefits to prices or wages, which grow more slowly than value added, whereas demand-responsive public services grow with the economy.

We also investigate the potential role of labor-augmenting AI that would increase worker productivity instead of replacing workers with capital. In this case, under demand-responsive provision, fiscal pressure rises when the proportional productivity gain is larger in the private sector, and falls when it is larger in the public-service sector. Finally, we discuss how heterogeneous labor or AI rents would modify the tax-base channel.

Our main results highlight the crucial role that public-service responses play in determining whether automation will ease fiscal pressure. One concrete implication is that, at unchanged tax rates and with other budget items unchanged as shares of value added, a reduction in the public-service share of value added is necessary for automation to ease fiscal pressure. Notably, public services can continue to grow in real terms, but not as a share of the economy. In the benchmark scenario for the EU-27, public services can grow by 7.3 percent without raising fiscal pressure under the existing tax system, compared with 11.3 percent under demand-responsive provision. We do not offer a prediction on how governments will respond.\footnote{The issue of optimal taxation in our setting is beyond the scope of this paper, and we do not rank fixed and demand-responsive provision in welfare terms. \citet{heathcote2021optimal} and \citet{ayaz2023who} consider optimal tax models without automation and study how the optimal income tax schedule should change when fiscal pressure rises. We also abstract from any increase in unemployment that AI may cause, and from interest payments on public debt, which lie outside the primary deficit we measure.} The growing consumer demand for public services may manifest in voters demanding substantial growth in public services, or as growth in privately financed provision. In the concluding section, we provide a general discussion of different possible margins of adjustment.

Our analysis brings together two broad literatures, one on Baumol's cost disease and one on automation and taxation. \citet{Baumol1967} expected municipal budgets to keep rising, because education, hospitals and similar services offer very limited scope for productivity growth. \citet{aghion2019artificial} apply this insight to AI and suggest that aggregate growth may be held back by sectors that are difficult to automate and that provide products that are ``essential'' (formally, gross complements to other goods). In our model, even when automation advances by the same amount in both sectors, the more labor-intensive public services instead become relatively more costly because of the economy-wide rise in wages.

\citet{andersen2016does} asks whether the cost disease makes tax-financed public services unsustainable using a model with a representative household paying a proportional income tax. At a fixed tax rate, public spending remains a constant share of GDP as manufacturing productivity rises.\footnote{Andersen also shows that the optimal tax rate remains constant under a particular preference specification, a result he describes as knife-edge.} \citet{andersen2017baumol} examine the same sustainability question with households that differ in ability. They show that adjustments to nonlinear income taxes can preserve public-service provision and every household's welfare. These adjustments raise average tax rates at every income level, whereas we hold tax rates fixed and measure how fiscal pressure changes. Moreover, because neither of these papers includes capital, income cannot shift between more and less heavily taxed factors. In our model, automation shifts income toward more lightly taxed capital, so fiscal pressure rises even when the public-service share stays constant.

Our production side builds on the task-based framework of \citet{acemoglu2018modeling, acemoglu2019automation}, and we treat AI as the latest form of automation.\footnote{Several studies measure the exposure of occupations and industries to AI \citep{FeltenRajSeamans2021, webb2019impact, eloundou2024gpts, engberg2024ai}. These measures capture technical potential, whereas actual use among US firms was still low in 2017 \citep{mcelheran2024ai}. We do not derive the size of our scenarios from them.} Using this framework, \citet{acemoglu2020does} argue that the US tax system is biased against labor and has thereby promoted inefficiently high levels of automation. We start from a similar gap between labor and capital taxes in the EU but treat automation as exogenous. Differences in this gap account for most of the variation across member states in how much fiscal pressure rises under demand-responsive provision in our benchmark scenario.

Part of this literature asks how taxes should be set as automation advances. These models treat taxes in detail but keep the expenditure side simple---they do not include public services with costs that could rise with growing wages. In the robot-tax model of \citet{guerreiro2022should}, government spending is a fixed fraction of consumption. \citet{thuemmel2023optimal} models how robots change the wages of different occupations, but treats revenue as an exogenous requirement. \citet{kleven_zidar_2026} show that automation calls for more progressive labor taxation, but their framework has no public-service sector. In the optimal-tax models of \citet{korinek_lockwood_2026}, the government's income finances transfers to households.

\citet{casas2024government} show in a calibrated model that, at existing tax rates, the ratio of tax receipts to output falls as automation advances, because the new technology replaces inputs that bear much of the tax burden, chiefly labor. \citet{bastani2024ai} discuss the broader observation that, with rates unchanged, faster growth could bring in more tax revenue and let governments spend more on publicly subsidized social services. Both are true in our model, where automation raises tax receipts through growth, but by less than value added. We add an expenditure side by modeling and quantifying how automation changes the cost of public services under different provision rules.

In sum, we contribute to the literature by providing the first quantitative assessment of how AI affects fiscal pressure in welfare states where labor-intensive public services are financed from taxes. We show that AI-driven automation can raise output \textit{and} tax revenue without easing fiscal pressure since the fiscal consequences of automation crucially depend on how public-service provision adjusts as the economy grows. Our quantitative results suggest that AI-driven automation is unlikely to ease fiscal pressure in the European context if public-service provision responds to consumer demand.

The rest of the paper proceeds as follows. Section~\ref{sec:model} sets out the model and Section~\ref{sec:cs} develops the comparative statics. Section~\ref{sec:cal} presents the calibration to the EU-27 and its member states, together with the quantitative results. Section~\ref{sec:ext} examines extensions, and Section~\ref{sec:concl} discusses policy responses and concludes.

\section{Model}\label{sec:model}

The model is a static two-sector general-equilibrium environment with task-based automation. A competitive market sector produces a numeraire private good, while a publicly financed service sector produces in-kind public services. Because the economy is small and open in capital markets, firms rent capital services at a world rental rate $R$ that is unaffected by domestic technology changes. This rental rate is gross of depreciation.\footnote{The capital stock responds to changes in technology, but we do not model household saving or how capital adjusts over time. Nor does the model have investment, so all value added is consumed, including the part that in practice replaces worn-out capital. Section~\ref{sec:capital_supply} reports results with the capital stock held fixed instead, and Section~\ref{sec:foreign_ownership} discusses foreign ownership.} Labor is fixed at $L=1$ and is perfectly mobile across domestic sectors but immobile internationally, so a common wage $W$ clears the domestic labor market. These assumptions focus the analysis on how automation reallocates a fixed workforce across sectors, draws additional capital into production at the given world rental rate, and shifts value added between labor and capital.

AI is modeled as an expansion of the set of tasks that capital can perform in place of labor. Except where noted, we assume that automation advances at least as much in market production as in public-service production. Since the workforce is fixed and the labor market clears, labor displacement refers to tasks that move to capital and not to a fall in headcounts.\footnote{We do not separately model the creation of new tasks for labor (the reinstatement channel of \citealt{acemoglu2019automation}). Instead, we interpret changes in the automation thresholds as net shifts from labor to capital after accounting for any such new tasks.}

\subsection{Production}\label{sec:prod}
The economy consists of a private market sector ($m$) and a public-service sector ($s$), or the private and public sectors for short. Within each sector, competitive task producers hire capital or labor to supply intermediate task outputs $y_j(i)$ indexed by $i\in[0,1]$, and a representative sectoral producer combines these into sectoral output $Y_j$. We assume that this producer uses a Cobb-Douglas aggregator,
\begin{equation}\label{eq:aggregator}
\log Y_j=\int_0^1\log y_j(i)\,di.
\end{equation}
This functional form makes every task essential with equal expenditure shares across tasks. Appendix~\ref{app:CES} generalizes the task aggregator to a constant elasticity of substitution (CES) form with elasticity $\sigma$ across task outputs.

Following the task-based automation framework of \citet{acemoglu2018modeling}, an exogenous automation threshold $I_j\in(0,1)$ for $j\in\{m,s\}$ determines which tasks can be performed by capital. Tasks with $i<I_j$ are automatable and can be performed by either capital or labor, while tasks with $i\geq I_j$ require labor:
\begin{equation}\label{eq:task_prod}
y_j(i)=
\begin{cases}
A_Kk_j(i)+l_j(i), & i<I_j,\\
l_j(i), & i\geq I_j,
\end{cases}
\end{equation}
where $k_j(i)$ and $l_j(i)$ are the capital and labor assigned to task $i$. In other words, capital and labor are perfect substitutes within automatable tasks.
We normalize labor productivity to one, and capital productivity $A_K>0$ is common to both sectors and held fixed when automation advances.\footnote{Appendix~\ref{app:factor_productivity} allows productivity to differ across sectors.}

\begin{assumption}[Production]\label{ass:initial}\leavevmode\\
\textit{(a) Automation saves cost.} The capital cost per unit of task output satisfies $R/A_K < 1$.

\noindent\textit{(b) Asymmetric sectoral automation.} Initially, the private sector is more automated than the public-service sector, $I_m > I_s$.
\end{assumption}

Producing one unit of task output costs $R/A_K$ with capital and $W$ with labor. Under the assumption $R/A_K < 1$, capital is cheaper than labor in automatable tasks, $R/A_K < W$.\footnote{\label{fn:capital_cheaper}If $W \leq R/A_K$, every task would have unit cost $W$, and so would the private good, whose price is one by choice of numeraire. Competition would then require $W = 1$, contradicting $W \leq R/A_K < 1$.} Firms therefore use capital below the threshold and labor above it, where capital is not available. As a consequence, the price $p_j(i)$ of task $i$ in sector $j$ is:
\begin{equation}\label{eq:task_prices}
p_j(i)=
\begin{cases}
R/A_K, & i<I_j,\\
W, & i\geq I_j.
\end{cases}
\end{equation}
At given factor prices, an increase in $I_j$ is cost-saving, since firms switch from labor to cheaper capital once a task becomes automatable. We take changes in $I_j$ as exogenous and abstract from adoption costs and gradual diffusion.

Under perfect competition, the cost of producing one unit of $Y_j$ must equal the minimum-cost combination of tasks that can deliver $Y_j$. The resulting sectoral unit cost $P_j$ is (see Appendix~\ref{app:task_equilibrium} for details):
\begin{equation}\label{eq:price_j}
P_j
\equiv
\exp\!\int_0^1\log p_j(i)\,di
=
\left(\frac{R}{A_K}\right)^{I_j}W^{1-I_j}.
\end{equation}

\noindent The first-order conditions imply that expenditure per task is equalized across tasks:
\begin{equation}\label{eq:per_task_expenditure}
p_j(i)y_j(i)=P_jY_j, \quad i\in[0,1].
\end{equation}
Since capital performs automatable tasks and labor performs other tasks, $y_j(i)=A_Kk_j(i)$ on $[0,I_j)$ and $y_j(i)=l_j(i)$ on $[I_j,1]$. Together with \eqref{eq:per_task_expenditure}, this implies that task output is constant within each interval:
\begin{equation}\label{eq:task_outputs}
y_j(i)=
\begin{cases}
A_KP_jY_j/R, & i<I_j,\\
P_jY_j/W, & i\geq I_j.
\end{cases}
\end{equation}
Integrating over $[0,I_j)$ and $[I_j,1]$ gives sectoral capital and labor demand:
\begin{equation}\label{eq:factor_demands}
\begin{aligned}
K_j &= \int_0^{I_j} k_j(i)\,di = \int_0^{I_j} \frac{y_j(i)}{A_K}\,di = \int_0^{I_j} \frac{P_jY_j}{R}\,di = \frac{I_jP_jY_j}{R}, \\
L_j &= \int_{I_j}^1 l_j(i)\,di = \int_{I_j}^1 y_j(i)\,di = \int_{I_j}^1 \frac{P_jY_j}{W}\,di = \frac{(1-I_j)P_jY_j}{W}.
\end{aligned}
\end{equation}
Multiplying \eqref{eq:factor_demands} by factor prices and dividing by total cost $P_jY_j$ gives the sectoral capital and labor cost shares:
\begin{equation}\label{eq:cost_shares}
s_K^j=I_j,
\qquad
s_L^j=1-I_j.
\end{equation}
Substituting the cost-minimizing task allocations into \eqref{eq:aggregator} yields sectoral production functions:\footnote{Starting from $\log Y_j = \int_0^1 \log y_j(i)\,di$ and using the piecewise-constant task outputs, the integral splits into two pieces with constant integrands,
\[
\log Y_j = \int_0^{I_j}\log\!\frac{A_KK_j}{I_j}\,di + \int_{I_j}^1\log\!\frac{L_j}{1-I_j}\,di = I_j\log\!\frac{A_KK_j}{I_j} + (1-I_j)\log\!\frac{L_j}{1-I_j}.
\]
Exponentiating both sides yields \eqref{eq:cobbdouglas}.}
\begin{equation}\label{eq:cobbdouglas}
Y_j=\mathcal{A}_jK_j^{I_j}L_j^{1-I_j},
\end{equation}
where $\mathcal{A}_j\equiv
\frac{A_K^{I_j}}{I_j^{I_j}(1-I_j)^{1-I_j}}$ is the effective sectoral TFP.

The private good is sold at a market price, whereas public services are provided in kind. We nevertheless use $P_s$ in \eqref{eq:price_j} to measure how many units of the private good could be produced for the cost of one unit of public services. This measure, which we call the unit resource cost of public services, treats the public-service producer as cost-minimizing at the same factor prices $W$ and $R$ as the private sector.\footnote{This abstracts from public--private wage gaps and procurement costs. Governments may also have wage-setting power, particularly in occupations where they are the main employer. If public wages rise less than private wages, the wage-driven increase in public-service costs would be smaller.} Since the cost shares in \eqref{eq:cost_shares} sum to one, payments to capital and labor add up to the cost of producing sectoral output:
\begin{equation}\label{eq:zeroprofit}
P_jY_j=RK_j+WL_j,\qquad j\in\{m,s\}.
\end{equation}

Recall that the private market good is the numeraire, so its producer price is $P_m=1$. From \eqref{eq:price_j}, this normalization pins down the equilibrium wage:
\begin{equation}\label{eq:wage}
W=\left(\frac{A_K}{R}\right)^{I_m/(1-I_m)}.
\end{equation}
Let $g\equiv I_m-I_s$ denote the threshold gap, which by \eqref{eq:cost_shares} also equals the public--private labor-intensity gap $s_L^s-s_L^m$. Combining \eqref{eq:price_j} with \eqref{eq:wage} then gives the unit resource cost of public services:
\begin{equation}\label{eq:Ps}
P_s=\left(\frac{A_KW}{R}\right)^{g}.
\end{equation}
Public services are more labor intensive than market production, so their unit cost rises with the cost of labor relative to capital, $A_KW/R$. The same ratio determines the cost saving from performing a task with capital instead of labor, $1-R/(A_KW)$.

\subsection{Aggregate output and resource accounting}\label{sec:resource}

Let $C_m$ denote household consumption of the private good and $C_s$ the quantity of public services provided to households, with $C_s$ set by the provision rules in Section~\ref{sec:hh} below. Market clearing in the private sector requires $Y_m=C_m$, and because public services are provided in kind, $Y_s=C_s$ holds by construction. Let $K\equiv K_m+K_s$ and $L\equiv L_m+L_s=1$ denote aggregate capital and labor. Gross value added, measured before consumption taxes and in units of the private good, is
\begin{equation}\label{eq:gdp_resource}
Y\equiv Y_m+P_sY_s=C_m+P_sC_s.
\end{equation}
Equivalently, summing \eqref{eq:zeroprofit} across sectors and using $P_m=1$ gives the income-side expression for $Y$:
\begin{equation}\label{eq:gdp_income}
Y=WL+RK.
\end{equation}

Dividing public-service spending, labor income and capital income by value added gives the economy-wide public-service expenditure share, labor share and capital share:
\begin{align}\label{eq:econ_shares}
s_E\equiv \frac{P_sC_s}{Y},\qquad
s_L\equiv \frac{WL}{Y},\qquad
s_K\equiv \frac{RK}{Y}.
\end{align}
Together with the income-side identity \eqref{eq:gdp_income}, the definitions in \eqref{eq:econ_shares} imply $s_L+s_K=1$. On the expenditure side, combining \eqref{eq:econ_shares} with \eqref{eq:gdp_resource} shows that $s_E$ is the public-service sector's share of value added and $1-s_E$ is the market sector's share. The economy-wide labor share is therefore the average of the sectoral labor cost shares from \eqref{eq:cost_shares}, weighted by these value-added shares:
\begin{equation}\label{eq:sL_identity}
s_L=s_L^m(1-s_E)+s_L^ss_E=(1-I_m)(1-s_E)+(1-I_s)s_E=1-I_m+g\,s_E.
\end{equation}
Holding the expenditure share fixed, an increase in $I_m$ or $I_s$ lowers $s_L$ by reducing labor's cost share in the sector where automation expands. In contrast, a rise in $s_E$ raises $s_L$ because it shifts value added toward the more labor-intensive public-service sector.

\subsection{Government and fiscal pressure}\label{sec:gov}

We now turn from production and resource accounting to the government's budget.
The government provides public services and finances them with linear taxes on labor and capital income at fixed effective rates $\tau_L$ and $\tau_K$, and with a tax on private consumption at a fixed rate $\tau_c\geq0$. The factor-income tax rates satisfy $0\leq\tau_K<\tau_L\leq1$. These taxes are levied on households, so they do not enter firms' cost minimization conditions. Labor-tax revenue is $\tau_LWL$ and capital-tax revenue is $\tau_K RK$, where $\tau_K$ is the effective tax rate on gross capital income.\footnote{If $\widetilde{\tau}_K$ is the tax rate on capital income net of depreciation, then $\tau_K=(1-\delta/R)\widetilde{\tau}_K$, where $\delta\in[0,R)$ is a depreciation allowance set by the tax code.} Private consumption costs households $1+\tau_c$ per unit and generates tax revenue $\tau_c C_m$, whereas public services are exempt from consumption tax.

We denote by $T$ the residual financing need for the modeled public-service bundle $C_s$:
\begin{equation}\label{eq:T}
T=P_sC_s-\tau_LWL-\tau_K RK-\tau_c C_m.
\end{equation}
$T$ can be read as the lump-sum tax, or transfer if negative, needed to balance the budget at the given tax rates. If the government instead borrows from households, or lends to them when $T<0$, then $T$ is the primary deficit of the modeled budget (the deficit excluding interest payments), and production and consumption are the same as with the tax. Dividing \eqref{eq:T} by $Y$ and using $s_K=1-s_L$ and $C_m/Y=1-s_E$ from \eqref{eq:gdp_resource} gives
\begin{equation}\label{eq:TY}
\begin{aligned}
\frac{T}{Y}&=s_E-\tau_Ls_L-\tau_K(1-s_L)-\tau_c(1-s_E) \\
&=(1+\tau_c)s_E-\tau_K-\tau_c-(\tau_L-\tau_K)\,s_L.
\end{aligned}
\end{equation}
The left-hand side, $T/Y$, is the residual financing need as a share of value added.\footnote{GDP at market prices also includes consumption-tax revenue and equals $Y+\tau_c C_m$.} We call it \emph{fiscal pressure}. Changes in fiscal pressure show how automation affects the primary deficit relative to value added. This effect works through public-service spending and the taxes in the model, and we measure it before the government adjusts taxes or spending to offset it. Other parts of the budget, such as cash transfers, can make this effect larger or smaller, as Section~\ref{sec:transfers_extension} illustrates.

The restriction $\tau_K<\tau_L$ reflects the empirically relevant case in which labor income is taxed more heavily than capital income, so the labor--capital tax differential $\tau_L-\tau_K$ is strictly positive. Since tax rates are fixed, differentiating \eqref{eq:TY} gives
\begin{equation}\label{eq:dTY}
d(T/Y)=(1+\tau_c)ds_E-(\tau_L-\tau_K)\,ds_L,
\end{equation}
where $ds_E$ and $ds_L$ are the total differentials of $s_E$ and $s_L$, which we use because both automation thresholds, $I_m$ and $I_s$, may change in the comparative statics.

Equation~\eqref{eq:dTY} separates the fiscal effect into two terms. The first, the expenditure contribution $(1+\tau_c)ds_E$, captures the change in public-service expenditure and the associated change in consumption-tax revenue. A rise in $s_E$ increases public expenditure and reduces the private-consumption tax base, both as shares of value added. The second, the tax-base contribution $-(\tau_L-\tau_K)ds_L$, shows how a lower labor share raises fiscal pressure because the labor--capital tax differential is positive. Because the economy remains at full employment by assumption, this channel works entirely through how labor and capital are allocated across sectors and through their shares in value added within each sector.

\subsection{Provision rules and households}\label{sec:hh}

Public services are provided in kind, so households do not buy $C_s$ at price $P_s$. Instead, $C_s$ is set by a provision rule, and we consider two cases.

\paragraph{Fixed provision.} The government commits to an institutionally fixed service bundle $C_s=\widehat{C}_s$ (school places, hospital treatments, care hours). Automation then affects the unit resource cost $P_s$ but leaves the quantity $C_s$ unchanged. Since $s_E=P_s\widehat{C}_s/Y$,
\begin{equation}\label{eq:dsE_fixed}
ds_E=s_E\bigl(d\log P_s-d\log Y\bigr).
\end{equation}
The expenditure share therefore falls whenever output grows faster than the unit cost of public services.

\paragraph{Demand-responsive provision.} The government lets $C_s$ track latent demand, the bundle households would choose if public services were priced at their unit resource cost $P_s$ and private consumption at $1+\tau_c$. We treat this as a reduced-form provision rule and do not derive it as the optimal rule from a planner's problem. For a given consumption-tax rate, we represent the rule by a smooth, strictly positive function for the ratio of public to private expenditure before consumption taxes,
\begin{equation}\label{eq:odds}
\frac{P_sC_s}{C_m}=\Omega(P_s),
\qquad \Omega:(0,\infty)\to(0,\infty).
\end{equation}
Because the ratio depends on $P_s$ alone, the level of income does not shift it at the given consumption-tax rate, as under homothetic preferences. The advantage of writing the rule as \eqref{eq:odds} is that it lets us summarize the response of the public-service expenditure share $s_E$ without deriving it from a full household problem.\footnote{Appendix~\ref{app:ces_eta} derives $\Omega$ for CES preferences over private and public consumption.} We define the elasticity of this ratio with respect to the public-service unit cost, evaluated at the initial $P_s$,
\begin{equation}\label{eq:etaP}
\eta_P\equiv
\frac{d\log \Omega(P_s)}{d\log P_s}.
\end{equation}
The resource identity \eqref{eq:gdp_resource} and the share definitions \eqref{eq:econ_shares} imply $s_E/(1-s_E)=P_sC_s/C_m$, which equals $\Omega(P_s)$ by \eqref{eq:odds}. Differentiating and using \eqref{eq:etaP} gives
\begin{equation}\label{eq:dsE_eta}
ds_E=s_E(1-s_E)\,\eta_P\,d\log P_s.
\end{equation}
At an interior equilibrium, $s_E(1-s_E)>0$, so the sign of $ds_E$ follows the sign of $\eta_P$ whenever $d\log P_s>0$. For a given initial expenditure share, the provision rule enters the local change in fiscal pressure only through $\eta_P$, so the local results below hold for any schedule with this local elasticity and do not rely on a specific preference specification. At a fixed consumption-tax rate, CES preferences, which correspond to a special case of \eqref{eq:odds}, imply $\eta_P=1-\varepsilon$, where $\varepsilon$ is the elasticity of substitution between private and public consumption.

\paragraph{Household budget.} For accounting completeness, we write the household-side counterpart to the government budget. We treat households as receiving labor income and the capital income generated by domestic production, paying factor and consumption taxes as well as the lump-sum residual $T$ (or receiving it if $T<0$):
\begin{equation}\label{eq:hhbudget}
WL(1-\tau_L)+RK(1-\tau_K)=(1+\tau_c)C_m+T.
\end{equation}
This simply states who receives factor income and how the government's residual financing need is paid. Preferences over private and public consumption enter only through the demand-responsive schedule $\Omega$ in \eqref{eq:odds}, and we do not conduct a welfare analysis. Our question is how automation changes the gap between public-service spending and tax revenue, and we leave aside how that gap is closed. In the model, $T$ closes it, although actual governments could respond in other ways.

The baseline model assigns all capital income to domestic residents, an assumption we relax in Section~\ref{sec:foreign_ownership}.

\subsection{Equilibrium}\label{sec:eq}

The preceding subsections set out technology and factor pricing, the public-service provision rule, and the government budget. We now collect these conditions into an equilibrium.

\begin{definition}[Equilibrium]\label{def:eq}
Given production and tax primitives $(I_m,I_s,R,A_K,\tau_L,\tau_K,\tau_c)$ together with a provision rule, either a fixed quantity $\widehat C_s$ or a demand-responsive schedule $\Omega$, an equilibrium consists of a wage $W$, a public-service unit cost $P_s$, sectoral allocations $\{L_j,K_j,Y_j\}_{j\in\{m,s\}}$, and a budget residual $T$ such that, with $C_m\equiv Y_m$, $C_s\equiv Y_s$, $L\equiv L_m+L_s$, and $K\equiv K_m+K_s$:
\begin{enumerate}
\renewcommand{\labelenumi}{\emph{\roman{enumi})}}
\item \emph{Cost minimization.} $W$ and $P_s$ satisfy \eqref{eq:wage}--\eqref{eq:Ps}, and factor demands are given by \eqref{eq:factor_demands}.
\item \emph{Factor allocation.} Total labor supply is fixed: $L=1$. Capital is supplied perfectly elastically at the exogenous rental rate $R$, so $K$ is whatever firms rent at this rate.
\item \emph{Provision rule.} Either $C_s=\widehat C_s$, or the expenditure ratio satisfies \eqref{eq:odds}.
\item \emph{Government budget.} $T$ is given by \eqref{eq:T}.
\end{enumerate}
We restrict attention to interior solutions, with $L_j,K_j>0$ for $j\in\{m,s\}$ and $C_m,C_s>0$ (so $s_E\in(0,1)$). Under fixed provision, $L_m>0$ requires $L_s<1$, which bounds the feasible range of the fixed quantity $\widehat C_s$.
\end{definition}

The equilibrium can be solved recursively. First, the production side pins down $W$ and $P_s$ from $(I_m,I_s,R,A_K)$ through \eqref{eq:wage}--\eqref{eq:Ps}. Given these prices, the provision rule sets the public-service quantity $C_s=\widehat C_s$ under fixed provision, or the expenditure share $s_E$ through \eqref{eq:odds} under demand-responsive provision. Labor-market clearing $L_m+L_s=1$ and the factor demands in \eqref{eq:factor_demands} then determine output and the sectoral allocation of labor and capital. Consumption follows as $C_m=Y_m$ and $C_s=Y_s$, and under demand-responsive provision the public-service quantity is $C_s=s_EY/P_s$. Finally, the budget residual $T$ is given by \eqref{eq:T}.\footnote{Combining \eqref{eq:hhbudget} with \eqref{eq:T} gives the goods-market identity $C_m+P_sC_s=WL+RK=Y$. Conversely, given this identity, \eqref{eq:T} implies \eqref{eq:hhbudget}. The household and government budgets are therefore not independent equilibrium restrictions. We use \eqref{eq:T} to pin down $T$, and \eqref{eq:hhbudget} then follows.}

Because the production side determines $W$ and $P_s$ independently of the provision rule, the cost of providing one unit of public services does not depend on the total quantity provided. For a given consumption-tax rate and provision rule, changes in factor-income tax rates affect only the residual financing need $T$, leaving wages, prices, production and consumption unchanged. This is because labor supply and the rental rate are fixed, factor-income taxes are levied on households, and $T$ adjusts as a lump-sum tax or transfer.

\section{Comparative statics}\label{sec:cs}

We organize the fiscal-pressure results around local changes in the automation thresholds $(I_m,I_s)$ under a given provision rule. The analysis first signs the responses of the wage and the public-service unit cost, then derives the fiscal-pressure decomposition and applies it to the two provision rules. The last subsection extends the decomposition to finite changes. Alternative production assumptions are discussed with the other extensions in Section~\ref{sec:ext}.

\subsection{Setup and production-side responses}\label{sec:setup_prices}

In the local analysis of this section, we consider automation changes satisfying
\[
dI_m\geq dI_s\geq 0,
\qquad dI_m>0.
\]

We work under Assumption~\ref{ass:initial} and restrict attention to interior equilibria as in Definition~\ref{def:eq}.

Differentiating the wage and public-service cost equations, \eqref{eq:wage} and \eqref{eq:Ps}, gives the following result.

\begin{result}[Wage and public-service cost responses]\label{prop:prices}
A local automation change raises the wage and the public-service unit cost,
\begin{align}
d\log W &= \frac{\log(A_KW/R)}{1-I_m}\,dI_m > 0, \label{eq:dlogW} \\
d\log P_s &= \frac{\log(A_KW/R)}{1-I_m}\left[g\,dI_s+(1-I_s)\,dg\right] > 0, \label{eq:dlogPs}
\end{align}
where $dg=dI_m-dI_s$.
\end{result}

The wage response is positive because $W>R/A_K$ and $dI_m>0$. Automating a market task lowers production costs at the initial wage. With the private-good price and the rental rate fixed, the wage rises until unit cost again equals that price. In \eqref{eq:dlogPs}, both terms in the bracket are nonnegative, and at least one is strictly positive because $g>0$ and $dI_s+dg=dI_m>0$. It should be noted that both responses rely on the elastic supply of capital. With a fixed capital stock, automation can raise the rental rate and lower the wage \citep{acemoglu2018race}, as it does in the calibrated economy of Section~\ref{sec:capital_supply}.

Result~\ref{prop:prices} captures the cost-disease channel of \citet{Baumol1967}, in which the public sector inherits the wage increase without enough automation of its own to offset it. Equation~\eqref{eq:dlogPs} shows that the public-service cost response depends both on the initial gap $g$ between public- and private-sector labor shares and on its change $dg$. In particular, equal percentage-point increases in automation, $dI_m=dI_s>0$, leave $g$ unchanged but still raise the relative cost of public services because the wage rises and public services remain more labor intensive.

\subsection{Fiscal-pressure decomposition}\label{sec:decomp}

With the responses of $W$ and $P_s$ signed, we now turn to the labor-share response $ds_L$ in the fiscal-pressure identity \eqref{eq:dTY}. Differentiating \eqref{eq:sL_identity}, $s_L=1-I_m+g\,s_E$, gives
\begin{equation}\label{eq:dsL_compact}
\begin{aligned}
ds_L=-dI_m+s_E(dI_m-dI_s)+g\,ds_E=-\underbrace{\bigl[(1-s_E)\,dI_m+s_E\,dI_s\bigr]}_{\equiv D}+g\,ds_E =-D+g\,ds_E.
\end{aligned}
\end{equation}
The term $D$ measures \emph{labor displacement}, the fall in the aggregate labor share that would occur if the sectors' shares of value added were held fixed ($ds_E=0$). Here $D>0$ because $dI_m>0$ and $dI_s\geq0$. The second term, $g\,ds_E$, works through the sector mix. A rise in $s_E$ shifts value added toward the public-service sector, which is more labor intensive since $g>0$, and so raises the aggregate labor share.

Substituting \eqref{eq:dsL_compact} into the fiscal-pressure identity \eqref{eq:dTY} gives the fiscal-pressure decomposition
\begin{equation}\label{eq:dTY_decomp}
d(T/Y)=\bigl[1+\tau_c-(\tau_L-\tau_K)g\bigr]\,ds_E+(\tau_L-\tau_K)\,D.
\end{equation}

We note two properties of \eqref{eq:dTY_decomp}. First, the net expenditure term $\bigl[1+\tau_c-(\tau_L-\tau_K)g\bigr]\,ds_E$ has the same sign as $ds_E$, because $g\in(0,1)$ and $0<\tau_L-\tau_K\leq1$ imply $\tau_c<1+\tau_c-(\tau_L-\tau_K)g<1+\tau_c$. Second, the displacement term $(\tau_L-\tau_K)D$ is strictly positive because $D>0$.

To understand the net expenditure term in \eqref{eq:dTY_decomp}, note that a change in $s_E$ affects both public expenditure and consumption-tax revenue, the two effects captured by the first term on the right-hand side of \eqref{eq:dTY}. However, the $g\,ds_E$ term in \eqref{eq:dsL_compact} shows that such a change also alters the labor-income tax base through its effect on the labor share, because public services are more labor intensive than market production. This means that when $s_E$ rises, value added shifts toward labor and factor-income tax revenue increases, offsetting part of the expenditure increase and the loss of consumption-tax revenue. In contrast, the displacement term in \eqref{eq:dTY_decomp} involves no expenditure effect. Labor displacement lowers the labor-income tax base relative to value added and therefore raises fiscal pressure.

A useful property of the decomposition in \eqref{eq:dTY_decomp} is that it can be written entirely in terms of a small set of sufficient statistics. Once $ds_E$, $D$, $g$, $\tau_L-\tau_K$, and $\tau_c$ are known, the fiscal-pressure response follows without tracking the full allocation of labor and capital across tasks. The gap $g$, the expenditure share $s_E$ and the tax rates can be measured in national accounts and tax data, whereas $D$ depends in addition on the size and sectoral split of the automation change. The response $ds_E$ also depends on capital's cost advantage in automated tasks, $\log(A_KW/R)$, and on the provision rule. The Cobb-Douglas task structure we have imposed makes the displacement term in \eqref{eq:dTY_decomp} especially transparent, because each additional automated task lowers the corresponding sectoral labor share one-for-one.

\begin{result}[Fiscal pressure and the expenditure-share response]\label{prop:fiscal_general}
Under the maintained assumptions and for the local automation changes considered, fiscal pressure increases whenever $ds_E\geq0$.
\end{result}

\begin{proof}
In \eqref{eq:dTY_decomp}, the coefficient $1+\tau_c-(\tau_L-\tau_K)g$ is positive and $(\tau_L-\tau_K)D>0$. Hence $ds_E\geq0$ implies $d(T/Y)>0$.
\end{proof}

\subsection{Fiscal pressure under fixed provision}\label{sec:fixed_provision}

Result~\ref{prop:prices} shows that automation raises the unit cost of public services. Under fixed provision, however, output grows faster than this cost (Appendix~\ref{app:fixed_share}), so \eqref{eq:dsE_fixed} gives $ds_E<0$. This is because capital performs part of the public-service tasks, so the cost of the fixed service bundle rises by less than the wage, and the workers it no longer needs move into market production, where value added rises by more than the wage. The labor share falls as well, since \eqref{eq:dsL_compact} gives $ds_L=-D+g\,ds_E<0$. The two terms in \eqref{eq:dTY} therefore have opposite signs, which makes the overall effect on fiscal pressure ambiguous. The lower expenditure share reduces fiscal pressure, while the lower labor share raises it because labor income is taxed more heavily than capital income. Since factor-income tax rates do not affect $ds_E$ or $ds_L$, the net effect depends on the tax differential, and \eqref{eq:dTY} shows that fiscal pressure rises if and only if $\tau_L-\tau_K$ exceeds $(1+\tau_c)\,ds_E/ds_L$, the ratio of the two share responses scaled by the consumption-tax factor. A larger $\tau_L-\tau_K$ gives more weight to the revenue lost as income shifts from labor to capital.

\subsection{Fiscal pressure under demand-responsive provision}\label{sec:provrules}

Under demand-responsive provision, the expenditure-share response in \eqref{eq:dsE_eta} has the same sign as $\eta_P$, because $s_E(1-s_E)>0$ and automation raises the unit cost of public services (Result~\ref{prop:prices}).

With CES preferences, as noted after \eqref{eq:dsE_eta}, $\eta_P=1-\varepsilon$. The cases $\eta_P>0$, $\eta_P=0$ and $\eta_P<0$ therefore correspond to gross complements ($\varepsilon<1$), the unit-elastic case ($\varepsilon=1$), and gross substitutes ($\varepsilon>1$) between private and public consumption.\footnote{The case $\eta_P=0$ should not be confused with fixed provision, which holds the real service quantity $C_s$ constant and yields $ds_E<0$ (Appendix~\ref{app:fixed_share}), whereas $\eta_P=0$ holds $s_E$ unchanged. Since $s_E=P_sC_s/Y$, a constant $s_E$ implies $d\log C_s=d\log Y-d\log P_s$.} With gross complements, the quantity of public services relative to private consumption falls by proportionally less than their relative cost rises, so spending shifts toward public services. This is the mechanism by which, in the structural-change model of \citet{ngai2007structural}, expenditure moves toward the sector with slower productivity growth.

\begin{result}[Fiscal pressure under demand-responsive provision]\label{prop:demand}
Under demand-responsive provision and for the local automation changes considered, fiscal pressure responds as follows.
\begin{enumerate}
\item[(i)] If $\eta_P\geq0$, fiscal pressure increases. Under CES preferences, this case corresponds to $\varepsilon\leq1$.
\item[(ii)] If $\eta_P<0$, both the expenditure share and the labor share fall, and the effect on fiscal pressure is ambiguous. Under CES preferences, this case corresponds to $\varepsilon>1$.
\end{enumerate}
\end{result}

\begin{proof}
Equation~\eqref{eq:dsE_eta} and Result~\ref{prop:prices} imply that $ds_E$ has the same sign as $\eta_P$. Case (i) follows from Result~\ref{prop:fiscal_general}. In case (ii), $ds_E<0$, so \eqref{eq:dsL_compact} gives $ds_L=-D+g\,ds_E<0$. In \eqref{eq:dTY_decomp}, the net expenditure term is then negative and the displacement term is positive. The first dominates when $\eta_P$ is sufficiently negative and the second when $\eta_P$ is close to zero, so the effect on fiscal pressure is ambiguous.
\end{proof}

When $\eta_P>0$ (gross complements under CES), the expenditure share rises and fiscal pressure increases even if reallocation toward public services raises the aggregate labor share. In this case, the labor share falls if and only if $D>g\,ds_E$, and when it does, both terms in \eqref{eq:dTY} raise fiscal pressure. In the unit-elastic case ($\eta_P=0$), the expenditure share is unchanged and fiscal pressure rises entirely through labor displacement, so that $d(T/Y)=(\tau_L-\tau_K)D>0$. Note that the positive sign under gross complements is not specific to automation. A private-sector labor-productivity gain also raises the wage and with it the unit cost of public services, and when $\eta_P>0$ it increases fiscal pressure through the net expenditure term (Appendix~\ref{app:factor_productivity}). What automation adds is the displacement term, which is positive even when the expenditure share is unchanged.

Result~\ref{prop:demand} also covers the case of equal percentage-point increases in automation in both sectors, $dI_m=dI_s=dI>0$. Labor displacement is then $D=dI$, and the public-service unit cost still rises because $g>0$. Fiscal pressure therefore increases whenever $\eta_P\geq0$, and in the unit-elastic case by $(\tau_L-\tau_K)\,dI$.

\subsection{Finite changes}\label{sec:finite_change}

The comparative statics derived above are local, whereas the quantitative analysis considers finite changes in automation. However, the fiscal-pressure identity carries over exactly, because \eqref{eq:TY} is linear in $s_E$ and $s_L$:
\begin{equation}\label{eq:finite_dTY_identity}
\Delta(T/Y)=(1+\tau_c)\Delta s_E-(\tau_L-\tau_K)\,\Delta s_L.
\end{equation}
The one complication is that the labor-share identity \eqref{eq:sL_identity}, $s_L=1-I_m+g\,s_E$, contains the product $g\,s_E$. Let $\Delta g=\Delta I_m-\Delta I_s$. Then
\[
\begin{aligned}
\Delta s_L
&=-\Delta I_m+\Delta(g\,s_E) \\
&=-\Delta I_m+g\,\Delta s_E+s_E\Delta g+\Delta g\,\Delta s_E \\
&=-\bigl[(1-s_E)\Delta I_m+s_E\Delta I_s\bigr]+(g+\Delta g)\Delta s_E.
\end{aligned}
\]
Writing
\[
D_\Delta\equiv (1-s_E)\Delta I_m+s_E\Delta I_s,
\qquad
g'\equiv g+\Delta g=g+\Delta I_m-\Delta I_s,
\]
with $s_E$ evaluated at the initial equilibrium, the exact change in the labor share is
\begin{equation}\label{eq:finite_dsL}
\Delta s_L=-D_\Delta+g'\Delta s_E.
\end{equation}
Equation~\eqref{eq:finite_dsL} uses the new threshold gap $g'$, whereas the local labor-share formula \eqref{eq:dsL_compact} uses the initial gap $g$. Substituting into \eqref{eq:finite_dTY_identity} gives
\begin{equation}\label{eq:finite_dTY_decomp}
\Delta(T/Y)=\bigl[1+\tau_c-(\tau_L-\tau_K)g'\bigr]\Delta s_E+(\tau_L-\tau_K)D_\Delta.
\end{equation}
Thus the finite decomposition \eqref{eq:finite_dTY_decomp} has the same form as the local decomposition \eqref{eq:dTY_decomp}, and the only adjustment needed when moving to finite changes is that $g'$ takes the place of $g$. Using $g$ instead would omit the second-order product $-(\tau_L-\tau_K)(\Delta I_m-\Delta I_s)\Delta s_E$. The condition for a falling labor share carries over in the same way, with $\Delta s_L<0$ exactly when $D_\Delta>g'\Delta s_E$.

We consider finite changes that raise automation in at least one sector and do not reduce it in either, provided that Assumption~\ref{ass:initial} still holds after the change and that each sector continues to produce using both labor and capital. Under these conditions, $D_\Delta>0$ and $0<g'<1$, so Result~\ref{prop:fiscal_general} applies to finite changes as well, with $d$ replaced by $\Delta$.

To calculate the finite change in fiscal pressure using \eqref{eq:finite_dTY_decomp}, we also need to know how $s_E$ changes under each provision rule. For this, we compute $W$ and $P_s$ before and after the automation change from \eqref{eq:wage} and \eqref{eq:Ps}.
Under fixed provision, $C_s$ is unchanged, so the finite version of \eqref{eq:dsE_fixed} is
\begin{equation}\label{eq:finite_sE_fixed}
\Delta\log s_E
=\Delta\log P_s-\Delta\log Y
=\Delta\log P_s-\Delta\log W+\Delta\log s_L.
\end{equation}
Here, a log change is the log of the new value minus the log of the initial value. The last equality uses $Y=W/s_L$, since $L=1$. Equations~\eqref{eq:finite_dsL} and \eqref{eq:finite_sE_fixed} together give the change in $s_E$ under fixed provision. In practice we obtain it in closed form by evaluating the expenditure ratio of Appendix~\ref{app:fixed_share} at the new thresholds. The argument of that appendix also applies to finite changes, because a higher wage or a higher public threshold lowers $P_s\widehat C_s/Y_m$ whatever the size of the change, so $\Delta s_E<0$ under fixed provision for every change considered here.

Under demand-responsive provision with CES preferences, the expenditure ratio is given by \eqref{eq:demand_share} in Appendix~\ref{app:ces_eta}. Taking log differences while holding preferences and the consumption-tax rate fixed gives
\begin{equation}\label{eq:finite_sE_demand}
\Delta\log\frac{s_E}{1-s_E}
=(1-\varepsilon)\Delta\log P_s.
\end{equation}
For a finite change with $\Delta I_m\geq\Delta I_s\geq0$ and $\Delta I_m>0$, \eqref{eq:wage} implies a higher wage. Since the gap between the thresholds does not shrink, the unit cost of public services in \eqref{eq:Ps} rises too. With $\varepsilon\leq1$, \eqref{eq:finite_sE_demand} then gives $\Delta s_E\geq0$, so case (i) of Result~\ref{prop:demand} carries over to finite changes under CES preferences.

\section{Calibration and quantitative results}\label{sec:cal}

We calibrate the model to the EU-27 and its member states using 2023 national accounts and tax data. These data supply the public-service share, sectoral labor shares, the rental rate and tax rates, which we combine with a common public--private substitution elasticity and an assumed 20 percent initial cost saving from performing a task with capital instead of labor. Calibration numbers are reported in Table~\ref{tab:eu_average}, where we also provide detailed sources. Table~\ref{tab:cross_country_calibration} in Appendix~\ref{app:countries} reports country-specific values.

\subsection{Parameters}
\paragraph{Sectors.} Using NACE industry codes, we assign public administration, defense, education, human health and social work to the public-service sector $s$ and all remaining industries to the private sector $m$.\footnote{The classification groups industries by activity, so publicly and privately operated hospitals belong to the same sector. This matches our assumption that the scope for automation depends on the tasks performed, not on whether the provider is public or private. Because these industries also include privately financed services, their value added only approximates tax-financed provision.} We measure the value of output at the sectoral ($P_jY_j$) and economy-wide ($Y$) levels by gross value added (the value of production minus intermediate inputs) since the model has no intermediate inputs. 

\paragraph{Automation thresholds.} With Cobb-Douglas task aggregation, the labor shares equal the shares of tasks performed by labor, $1-I_j$, so the automation threshold is $I_j=1-s_L^j$. We measure labor shares as employee compensation divided by gross value added. The capital share is one minus the labor share, $s_K^j=1-s_L^j=I_j$.\footnote{This capital share is gross of depreciation and includes operating surplus and the mixed income of the self-employed. In the private sector, operating surplus also includes the income attributed to owner-occupied housing, which raises measured $I_m$ without any increase in task automation. National accounts value non-market government output at cost, with no net return on government-owned capital, so measured $I_s$ leaves out any such return.} For the EU-27, the labor share is 0.461 in the private sector and 0.776 in the public-service sector, which gives $I_m=0.539$ and $I_s=0.224$. In the calibrated model, capital thus performs about 54 percent of private-sector tasks and 22 percent of public-service tasks.

Income of the self-employed, which the national accounts record as mixed (capital--labor) income, is treated as capital income in our calibration. Measured $I_m$ is therefore higher in countries with substantial self-employment, affecting in particular Greece, Italy, Poland, Romania and Slovakia. In our calibration, a higher $I_m$ amplifies the response of wages (and thus the unit cost of public services) to private-sector automation. Section~\ref{sec:cross_country} reports results under an alternative treatment of mixed income.

\paragraph{Public-service share.} We measure the public-service expenditure share, $s_E=P_sC_s/Y$, as the public-service sector's share of gross value added. It is 0.183 for the EU-27. This value defines the fixed quantity $\widehat C_s$ under fixed provision. It also sets the initial level of the schedule $\Omega$ at $P_s$ under demand-responsive provision. Both rules therefore start from the same equilibrium. The share excludes cash transfers, including pensions, because public services in the model are provided in kind. 

\paragraph{Labor share.} The economy-wide labor share is the average of the two sectoral labor shares, weighted by value added, $s_L=(1-s_E)(1-I_m)+s_E(1-I_s)=1-I_m+g\,s_E$, where $g=I_m-I_s$ is the gap between the thresholds. For the EU-27 this gives $s_L\simeq0.519$, which by construction equals the share of value added paid as employee compensation in the data.\footnote{The EU ratio of government consumption and investment to GDP was 0.246 in 2023 \citep{EurostatGov10aMain}. It is higher than our measure because it includes government investment and the intermediate inputs that government buys from market producers, which in the model are private-sector output. Using this ratio for $s_E$ would give $s_L\simeq0.538$, about two percentage points above the share of value added paid to employees in the data.}

\paragraph{Cost of capital.} The rental rate of capital, $R$, and capital productivity, $A_K$, affect our results only through their ratio $R/A_K$, the effective cost of capital. We therefore calibrate this ratio instead of estimating $A_K$ directly. We assume that replacing labor with capital in an automatable task initially saves 20 percent of the task's cost, $1-R/(A_KW)=0.2$.\footnote{The 20 percent saving is motivated by the task-level cost savings discussed by \citet{acemoglu2025simple}, who puts the average labor-cost saving in tasks affected by AI at about 27 percent. That is the figure to compare with ours, since in our model the cost of a task performed by labor is entirely labor cost. The underlying experiments study AI that assists workers, but we assume that savings of a similar size apply when AI performs a task in place of a worker. Appendix~\ref{app:saving_sensitivity} reports results for a 30 percent initial saving.} Equation~\eqref{eq:wage} implies $R/(A_KW)=(R/A_K)^{1/(1-I_m)}$, so the calibration sets $R/A_K=0.8^{1-I_m}$. The ratio varies across countries with $I_m$. We hold $A_K$ fixed as automation increases, so the cost saving rises above its initial 20 percent as wages grow.

To give $A_K$ a value, we compute the gross rental rate $R$ as capital income divided by the net capital stock in 2023.\footnote{For $R$, we measure capital income as GDP minus labor income, using AMECO's adjusted wage share, which treats each self-employed worker as earning the average compensation of an employee. This measure differs from gross value added minus compensation of employees in two ways. Using GDP instead of gross value added adds taxes on products net of subsidies, while imputing wages to the self-employed reduces measured capital income. On balance, the EU-27 value of $R$ is somewhat higher than it would be with gross value added minus compensation of employees. These differences do not affect the reported fiscal effects, because recalibrating $A_K$ to preserve the 20 percent initial saving keeps $R/A_K$ unchanged.} This gives $R=0.158$ for the EU-27, and we use this value for all countries. Dividing $R$ by the calibrated ratio $R/A_K$ gives $A_K=0.158/0.8^{0.461}\simeq0.175$ for the EU-27. Country values range from 0.166 to 0.178.

\paragraph{Substitution elasticity.} The public--private substitution elasticity governs the expenditure channel under demand-responsive provision. We set $\varepsilon=0.67$, the midpoint of the range of 0.60--0.74 that \citet{francois2023intratemporal} regard as their most reliable pooled estimates for 17 European countries over 1970--2018. Every estimate in this range is below one. Public and private consumption are therefore gross complements, and the expenditure share rises when public services become relatively more expensive, as in case (i) of Result~\ref{prop:demand}. We apply the same value to all countries.\footnote{The study also reports estimates for individual countries (its Table 8, common correlated effects estimator), which range from slightly below zero for Portugal to about 1.2 for Ireland. Those for Belgium, Portugal and Switzerland are not statistically different from zero.}

\paragraph{Tax rates.} The labor-tax rate $\tau_L$ is the implicit tax rate on labor, which equals taxes and social contributions on employed labor income divided by compensation of employees plus payroll taxes (0.370 for the EU-27).\footnote{Payroll taxes are levies on the wage bill or employment that the national accounts record as taxes on production and exclude from compensation of employees.} The capital-tax rate $\tau_K$ is capital-tax revenue divided by capital income, measured as gross value added minus compensation of employees (0.196 for the EU-27).\footnote{This rate only approximates the model's tax on capital income. Besides mixed income, measured capital income includes payroll taxes and other taxes on production net of other subsidies on production. Capital-tax revenue also contains taxes on self-employment income and on the stock of capital, such as property and wealth taxes. Our labor shares leave out payroll taxes, although the denominator of $\tau_L$ includes them. Section~\ref{sec:cross_country} examines alternative treatments of payroll taxes, taxes on the stock of capital and mixed income.} The labor--capital tax differential is thus 17.4 percentage points (pp) for the EU-27. This tax differential, ranging from 7pp in the Netherlands and Luxembourg to 29pp in Slovakia, determines the tax-base contribution for a given change in the labor share. 

The implicit tax rate on consumption divides consumption-tax revenue by spending including the tax, whereas the model applies $\tau_c$ to spending before tax. We convert each country's rate to the model's basis, which gives $\tau_c=0.196$ for the EU-27.\footnote{We treat the EU-27 implicit rate of 0.164 as 16.4 in taxes out of every 100 spent, leaving 83.6 before tax, so that $\tau_c=16.4/83.6\simeq0.196$. Applying this rate in the model is an approximation, because the model taxes all market-sector value added, a broader base than in the statistics.}

\begin{table}[tbp]
\centering\small
\caption{Calibrated parameters, EU-27, 2023}
\label{tab:eu_average}
\begin{tabularx}{\textwidth}{@{}l D c c S@{}}
\toprule
Parameter & Description & \begin{tabular}[b]{@{}c@{}}Cross-country\\range\end{tabular} & EU-27 & Source \\
\midrule
$R$ & Gross rental rate of capital & -- & 0.158 & AMECO \\
$A_K$ & Capital productivity, set for a 20 percent initial cost saving & [0.166, 0.178] & 0.175 & Calibrated \\
$\varepsilon$ & Public--private substitution elasticity & -- & 0.670 & \citet{francois2023intratemporal}\\
$s_E$ & Public-service share of gross value added & [0.096, 0.216] & 0.183 & Eurostat nama\_10\_a10 \\
$I_m$ & Private-sector automation threshold, $1-s_L^m$ & [0.471, 0.770] & 0.539  & Eurostat nama\_10\_a10\\
$I_s$ & Public-service automation threshold, $1-s_L^s$ & [0.116, 0.302] & 0.224 & Eurostat nama\_10\_a10 \\
$\tau_L$ & Implicit tax rate on labor income & [0.239, 0.442] & 0.370 & DG TAXUD Taxation Trends\\
$\tau_K$ & Tax rate on capital income & [0.074, 0.274] & 0.196 & DG TAXUD, Eurostat nama\_10\_a10\\
$\tau_c$ & Tax rate on consumption, producer-price basis & [0.156, 0.287] & 0.196 & DG TAXUD Taxation Trends, converted\\

\bottomrule
\end{tabularx}
\begin{flushleft}
\scriptsize
Notes: EU-27 values are for the EU as a whole, so they are not simple averages of the country values in Table~\ref{tab:cross_country_calibration}. $R$ and $\varepsilon$ are common to all countries, and the other parameters are country-specific. Observed rental rates across member states range from 0.128 to 0.409. For $\varepsilon$, the pooled estimates that \citet{francois2023intratemporal} regard as most reliable range from 0.60 to 0.74. The EU-27 values of $\tau_K$ and $\tau_c$ coincide by chance. They are measured from different data and differ in every member state.

Sources: National accounts and tax data refer to 2023. $s_E$, $I_m$, $I_s$ and the denominator of $\tau_K$ use gross value added (B1G) and compensation of employees (D1) from Eurostat table nama\_10\_a10 \citeyearpar{EurostatNama10a10}, extracted on 16 June 2026 as percentages of GDP rounded to one decimal. The public-service sector is NACE sections O--Q, and the private sector is all other activities. $\tau_L$ and $\tau_c$ are based on DG TAXUD's implicit tax rates on labor and consumption, and $\tau_K$ uses its taxes on capital as a percentage of GDP, from Tables 82, 81 and 53 of the Taxation Trends data \citep{ecTaxationTrends2026}. $R$ uses the adjusted wage share, GDP at 2020 reference levels and net capital stock at 2020 prices from AMECO \citep{ecAMECO2026}, extracted on 12 June 2026. Calculations for the benchmark calibration use the three-decimal inputs reported in Tables~\ref{tab:eu_average} and~\ref{tab:cross_country_calibration}. $A_K$ is calculated from $R$, $I_m$ and the 20 percent initial saving without further rounding.
\end{flushleft}
\end{table}

\subsection{The size of the automation shock}
Our intention is to illustrate the consequences of different scenarios that have been discussed in the AI literature. These scenarios should not be interpreted as forecasts. Instead, we show how fiscal pressure responds to automation of different magnitudes and sectoral distributions.

Historical changes in the labor share provide one possible benchmark for the size of the shocks since $s_L^j=1-I_j$. The expression implies that a 1pp increase in $I_j$ is associated with a 1pp decline in the sector's labor share. \citet{karabarbounis2014global} document a decline of about 5pp in the global corporate labor share over 1975--2012. The \citet{IMF2017LaborShare} also reports declines in advanced economies since the 1980s and in many emerging economies since the early 1990s, and \citet{bergholt2026dp21436} and \citet{Dao2017Why} examine the contribution of automation. These changes amount to a fall of roughly 0.1pp to 0.2pp per year.

AI could, however, automate a broader range of tasks than earlier technologies, and do so more quickly. \citet{karnofsky2016some} and \citet{brynjolfsson2025research} discuss this possibility as transformative AI. We therefore consider increases of 1pp, 5pp and 10pp in the private-sector automation threshold. A 1pp shock is roughly comparable to five to ten years of historical change, whereas a 10pp shock is larger than the entire decline described above. Even when confined to the private sector, it would lower the EU aggregate labor share by about 8pp under demand-responsive provision and almost 10pp under fixed provision. 

Our scenarios combine private-sector increases in the range of 1pp to 10pp with public-sector increases of zero, half, or all of the private-sector increase. We refer to a scenario as \textit{private-led} when private-sector automation increases more than public-sector automation, and \textit{symmetric} when the two increase by the same amount. Our \textit{benchmark scenario} raises the private-sector threshold by 5pp and the public-sector threshold by half as much (2.5pp).

\subsection{Quantitative results for the EU-27}\label{sec:quant}
We calculate the exact change in fiscal pressure after each automation shock using the finite-change expressions in Section~\ref{sec:finite_change}, and derive the expenditure and tax-base contributions using \eqref{eq:finite_dTY_identity}. We first show results for the EU-27, where we vary the size of the shock and its distribution across the two sectors to illustrate different scenarios. Section~\ref{sec:cross_country} then applies the benchmark shock to each member state, with and without its public-sector part. We use demand-responsive provision as the main case, and let fixed provision serve as a comparison.

\paragraph{Fiscal pressure.}
Table~\ref{tab:scenario_grid} reports the EU-27 results. For each provision rule, the first two columns split the change in fiscal pressure into its two sources. The \textit{expenditure contribution}, $(1+\tau_c)\Delta s_E$, is the change in public-service spending plus the consumption-tax revenue lost when the private share falls by the same $\Delta s_E$. The \textit{tax-base contribution}, $-(\tau_L-\tau_K)\Delta s_L$, is the revenue lost because income shifts from labor to more lightly taxed capital. Since all contributions are measured as shares of value added, ``lost revenue'' means revenue that grows more slowly than value added. Their sum is the change in fiscal pressure, $\Delta(T/Y)$, in percentage points of value added. The last column under each rule gives the growth of real value added in percent, to illustrate the size of the shock in a familiar metric.

\begin{table}[tbp]
\centering
\caption{Change in fiscal pressure by automation scenario, EU-27, 2023}
\label{tab:scenario_grid}
\footnotesize
\setlength{\tabcolsep}{3pt}\renewcommand{\arraystretch}{1.15}
\begin{tabular}{@{}l rr rrrr rrrr@{}}
\toprule
& & & \multicolumn{4}{c}{\textbf{Demand-responsive ($\boldsymbol{C_s}$)}} & \multicolumn{4}{c}{Fixed provision ($\widehat{C}_s$)} \\
\cmidrule(lr){4-7}\cmidrule(lr){8-11}
Scenario & $\Delta I_m$ & $\Delta I_s$ & Expenditure & \begin{tabular}[b]{@{}c@{}}Tax\\base\end{tabular} & $\boldsymbol{\Delta(T/Y)}$ & \begin{tabular}[b]{@{}c@{}}Output\\growth\\(\%)\end{tabular} & Expenditure & \begin{tabular}[b]{@{}c@{}}Tax\\base\end{tabular} & $\boldsymbol{\Delta(T/Y)}$ & \begin{tabular}[b]{@{}c@{}}Output\\growth\\(\%)\end{tabular} \\
\midrule
Private 1, Public 0 & 1 & 0 & 0.02 & 0.14 & 0.16 & 2.0 & $-$0.42 & 0.16 & $-$0.25 & 2.3 \\
Private 1, Public 0.5 & 1 & 0.5 & 0.02 & 0.16 & 0.17 & 2.2 & $-$0.49 & 0.18 & $-$0.31 & 2.5 \\
Symmetric 1 & 1 & 1 & 0.01 & 0.17 & 0.18 & 2.4 & $-$0.56 & 0.20 & $-$0.36 & 2.7 \\
Private 5, Public 0 & 5 & 0 & 0.12 & 0.70 & 0.83 & 11.0 & $-$2.12 & 0.82 & $-$1.29 & 12.7 \\
Private 5, Public 2.5 & 5 & 2.5 & 0.09 & 0.79 & 0.87 & 12.3 & $-$2.45 & 0.91 & $-$1.54 & 14.0 \\
Symmetric 5 & 5 & 5 & 0.05 & 0.87 & 0.92 & 13.5 & $-$2.78 & 1.00 & $-$1.78 & 15.3 \\
Private 10, Public 0 & 10 & 0 & 0.28 & 1.40 & 1.69 & 24.9 & $-$4.33 & 1.68 & $-$2.64 & 29.7 \\
Private 10, Public 5 & 10 & 5 & 0.20 & 1.57 & 1.77 & 28.0 & $-$4.95 & 1.84 & $-$3.11 & 33.0 \\
Symmetric 10 & 10 & 10 & 0.12 & 1.73 & 1.85 & 31.2 & $-$5.54 & 1.99 & $-$3.55 & 36.1 \\
\bottomrule
\end{tabular}
\begin{flushleft}
\scriptsize
Notes: Exact finite changes around the initial EU-27 equilibrium for 2023, with parameters as in Table~\ref{tab:eu_average} and $A_K$ held fixed at the value calibrated to a 20 percent initial cost saving ($A_K=0.175$). $T/Y$ is the residual financing need as a share of value added, so a positive change is a rise in fiscal pressure. For each provision rule, the expenditure column is the expenditure contribution $(1+\tau_c)\Delta s_E$, and the tax-base column is the tax-base contribution $-(\tau_L-\tau_K)\Delta s_L$. They sum to $\Delta(T/Y)$ by \eqref{eq:finite_dTY_identity}. $\Delta I_m$ and $\Delta I_s$ are in percentage points of the automation thresholds, so 5 raises $I_j$ by 0.05. Fiscal columns are in percentage points of gross value added. Components may not sum to the total because of rounding. Output growth is the percentage change in real value added between the two equilibria, measured by a Fisher quantity index that values the private good at its producer price and public services at their unit cost. Growth in $Y$, the denominator of the fiscal columns, also reflects changes in the unit cost of public services relative to the private good.
\end{flushleft}
\end{table}

The table shows that automation leads to increased fiscal pressure in \textit{every} demand-responsive scenario. Strikingly, fiscal pressure increases even in the symmetric cases where automation is as fast in the public as in the private sector. In fact, the adverse effect even survives in scenarios where the public sector automates \textit{faster} than the private sector (see Table~\ref{tab:scenario_grid_publicled} in Appendix~\ref{app:saving_sensitivity}). In sharp contrast, automation leads to falling fiscal pressure in \textit{every} fixed-provision scenario. The sign reversal is complete across the wide grid we explore, which illustrates the fundamental importance of endogenous provision of public services for understanding how automation affects fiscal pressure.

Furthermore, the magnitudes are substantial. In the benchmark scenario, fiscal pressure rises by 0.87pp of value added under demand-responsive provision and falls by 1.54pp under fixed provision. The demand-responsive increase of 0.87pp of value added (about 0.79pp of GDP) is about two-thirds of the rise of 1.2pp of GDP in EU age-related public spending that the 2024 Ageing Report projects for 2022--2070 \citep{EC2024AgeingReport}. In the symmetric 10pp scenario, where public-sector automation keeps pace with a private sector that automates at a transformative rate, fiscal pressure rises by even more (1.85pp of value added, or about 1.7pp of GDP) than in the Ageing Report scenario. 

These results are very robust to variations in the calibration parameters. Appendix~\ref{app:saving_sensitivity} explores scenarios with a 30 percent initial cost saving, an $\varepsilon$ between 0.60 and 0.74 or a consumption-tax rate of 0.142, and the conclusions are unchanged. Excluding taxes on the stock of capital, such as property and wealth taxes, leads to larger fiscal pressure but leaves qualitative results unchanged.

\paragraph{The tax-base and expenditure channels.}
The decomposition in Table~\ref{tab:scenario_grid} shows that almost all of the demand-responsive fiscal pressure comes from the \textit{tax-base} channel. Across all nine combinations we consider, the tax-base channel accounts for 83 to 95 percent of the total. In the benchmark scenario, 0.79pp of the 0.87pp increase comes from the tax base and only 0.09pp from expenditure. 

To understand the dominance of the tax-base channel, note first that equation~\eqref{eq:finite_dsL} separates the labor-share change into direct displacement and sectoral reallocation (the shift of value added toward the more labor-intensive public sector as $s_E$ rises). Direct displacement, $D_\Delta=(1-s_E)\Delta I_m+s_E\Delta I_s$, is about 4.5pp in the benchmark scenario, which multiplied by the labor--capital tax differential (0.174) gives a 0.79pp contribution to fiscal pressure.\footnote{Because Cobb-Douglas task aggregation lowers the sectoral labor share one-for-one with the automation threshold, this displacement effect depends only on the shocks, the initial expenditure share and the tax differential.} Sectoral reallocation partly restores the labor share, but the shift is too small to matter. The reported tax-base contribution is therefore close to the displacement term.\footnote{The tax-base contribution reported in Table~\ref{tab:scenario_grid} equals $-(\tau_L-\tau_K)\Delta s_L$, which combines the displacement term $(\tau_L-\tau_K)D_\Delta$ and the reallocation effect. The alternative decomposition of equation~\eqref{eq:finite_dTY_decomp} places the reallocation effect in the net expenditure term instead. }

The expenditure contribution is small. In the benchmark scenario, wages go up by 2.8 percent and the unit cost of public services rises by 1.5 percent. Evaluated at the initial equilibrium, \eqref{eq:dsE_eta} scales the cost increase by $s_E(1-s_E)\eta_P\approx 0.05$, so the expenditure share increases by less than 0.1pp. By our assumptions, demand is homothetic, so the expenditure share can only change through the cost channel, not through income growth, even though the quantity of public services $C_s$ rises by about 11 percent in the benchmark scenario. Our homothetic benchmark is conservative in this respect. If demand for public services instead rose more than in proportion to income, the expenditure contribution would be larger.\footnote{Some evidence points in this direction. The share of health spending in the United States has risen as income has grown, which \citet{hall2007value} explain by an income elasticity above one, and the share of services in consumption rises with income across countries \citep{herrendorf2014growth}.}

\paragraph{Value added and factor incomes.}
Real value added grows by 2 to 31 percent across the demand-responsive scenarios of Table~\ref{tab:scenario_grid} and by up to 36 percent under fixed provision. The value-added responses are larger under fixed provision because it shifts more labor into market production. Capital is supplied perfectly elastically at the given rental rate in our baseline model, so the capital stock responds strongly: the symmetric 10pp shock raises the capital stock by 59 percent under demand-responsive provision and 69 percent under fixed provision. Because labor supply and tax rates are fixed, wages, labor income and labor-tax receipts rise together in every scenario (2.8 percent in the benchmark). In the benchmark scenario, capital income grows by 23 percent under demand-responsive provision and 27 percent under fixed provision. With capital taxed at about 20 percent and labor at 37 percent, this shift of income toward capital causes tax receipts to grow more slowly than value added. For the EU as a whole, holding the rental rate fixed may overstate the responses of capital and output, because an automation shock common to many economies could raise the world rental rate. The joint assumption of a domestically owned perfectly elastic capital stock is important and we therefore explore the case of foreign-owned capital in Section~\ref{sec:foreign_ownership} and the case of a fixed capital stock (as a polar alternative to perfectly elastic capital) in Section~\ref{sec:capital_supply}.

\begin{figure}[tbp]
\centering
\includegraphics[width=\textwidth]{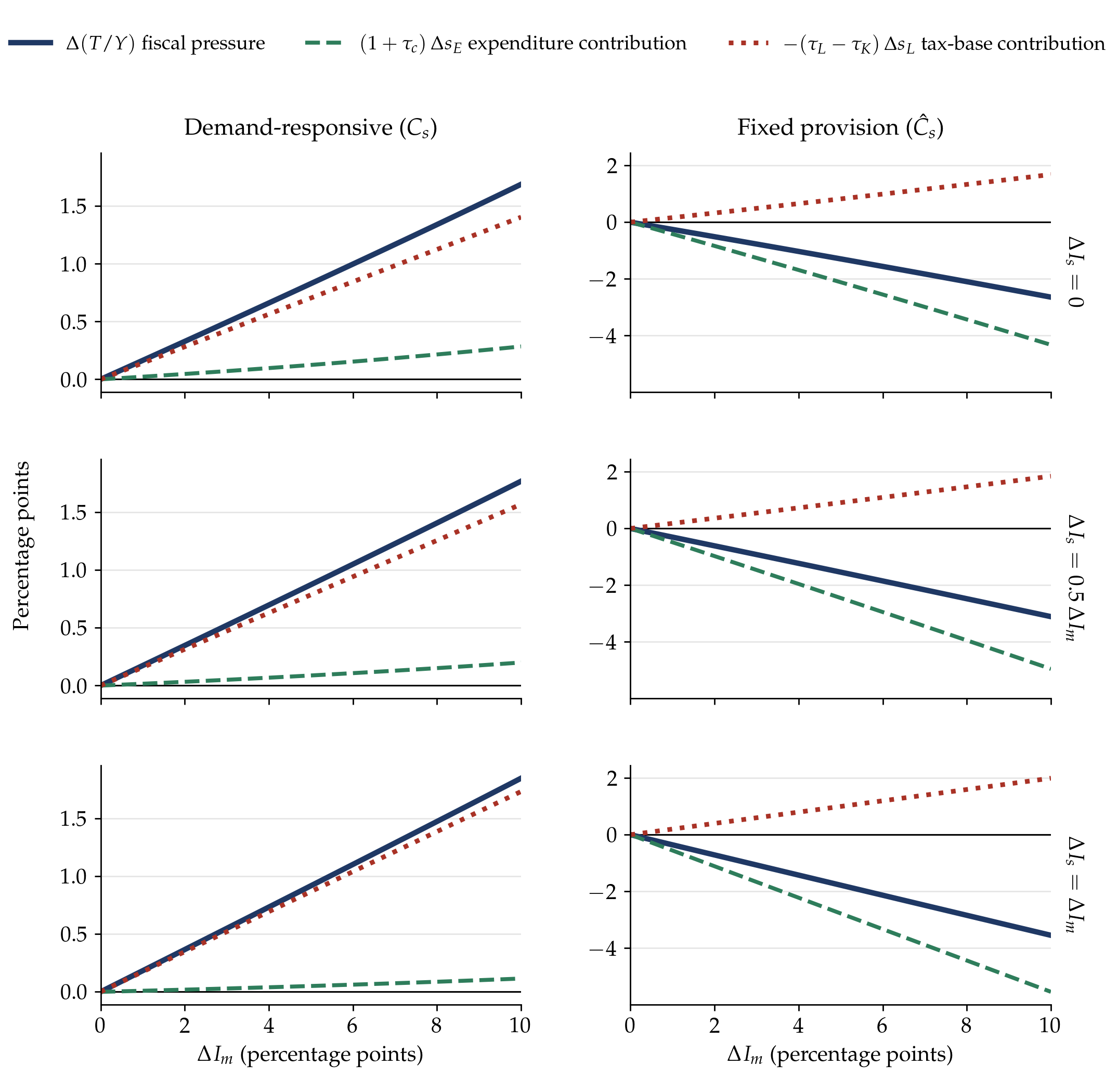}
\caption{Change in fiscal pressure and its two contributions, EU-27, 2023}
\label{fig:scenario_plots}
\begin{flushleft}
\scriptsize
Notes: Exact finite changes in percentage points of value added as $\Delta I_m$ varies from 0 to 10pp, with $\Delta I_s$ equal to zero, one half and all of $\Delta I_m$. The expenditure contribution $(1+\tau_c)\Delta s_E$ and the tax-base contribution $-(\tau_L-\tau_K)\Delta s_L$ add up to $\Delta(T/Y)$. Parameters as in Table~\ref{tab:eu_average}, with $A_K$ set for a 20 percent initial cost saving. Vertical scales differ across columns.
\end{flushleft}
\end{figure}

\paragraph{Public-sector automation.}
Figure~\ref{fig:scenario_plots} plots the two channels separately as the private-sector shock varies from 0 to 10pp, with public-sector automation set to zero, half, and all of the private-sector shock. It shows that, in the EU-27 calibration, fiscal pressure increases further under demand-responsive provision when public-sector automation catches up to private-sector automation. In the 5pp scenario, the fiscal effect rises from 0.83pp to 0.92pp as public automation goes from zero to matching the private shock. At the 10pp shock it rises from 1.69pp to 1.85pp. This result may appear counterintuitive, since public-sector automation lowers service production costs. Under fixed provision, these lower costs do indeed reduce fiscal pressure: the fall deepens from 1.29pp to 1.78pp in the 5pp scenario.

The reason for the counterintuitive result is that public-sector automation adds to the growth of value added (from 11 to 13.5 percent in the 5pp scenario), and under demand-responsive provision this growth raises the demand for public services. Their share of value added therefore falls only slightly compared with the case without public-sector automation, even though the Baumol effect is weakened as growth becomes more symmetric (the expenditure contribution falls from 0.12pp to 0.05pp). But for a given private-sector shock, additional public-sector automation leaves wages and total labor income unchanged, so all of the additional income accrues to capital, which is taxed at a lower rate. Tax revenues therefore fall further behind value added (the tax-base contribution to fiscal pressure rises from 0.70pp to 0.87pp), and this outweighs the lower production costs and the reduced Baumol effect. In part, this is because the Baumol effect is small to begin with. In Appendix~\ref{app:saving_sensitivity} we provide three ``public-led'' scenarios where automation is faster in the public sector. There, falling public-service unit costs make the expenditure contribution negative, but the effect on fiscal pressure is positive even in these scenarios.\footnote{Result~\ref{prop:demand} assumes that automation advances at least as much in the market sector, so the result does not cover these scenarios and the sign of their fiscal effect depends on the calibrated parameters.}

\paragraph{Fixed provision.}
Under fixed provision, output grows faster than the unit cost of public services, so the public-service expenditure share falls, causing fiscal pressure to drop. Figure~\ref{fig:scenario_plots} shows that this pattern holds across the full scenario grid. The expenditure contribution falls sharply with private-sector automation, and the tax-base contribution rises, but the sum remains negative throughout. In the benchmark scenario of Table~\ref{tab:scenario_grid}, real value added grows by 14 percent, the expenditure contribution is $-2.45$pp and the tax-base contribution is $+0.91$pp, resulting in a net fall in fiscal pressure of 1.54pp. Additional public-sector automation lowers the cost of the fixed bundle and frees public-sector workers for market production, so output rises by more and the fiscal improvement is larger.

The fixed public-service bundle implies that value added shifts toward the less labor-intensive private sector as the economy grows. The labor share therefore falls by more under fixed provision than under demand-responsive provision. As a consequence, the tax-base contribution is larger (0.91pp) than its demand-responsive counterpart (0.79pp).

A notable aspect of the fixed-provision scenario is that consumption shares change as the economy grows through automation. These effects may be quite substantial: in the benchmark scenario, public services remain constant while private consumption grows by 17 percent.

\paragraph{Provision that leaves fiscal pressure unchanged.}
A natural follow-up to the two polar cases is to ask what level of public-service provision would leave fiscal pressure unchanged. In the benchmark scenario, public services can grow by 7.3 percent without raising fiscal pressure, compared with 11.3 percent under demand-responsive provision and no growth by design under fixed provision. Appendix~\ref{app:neutral_provision} solves \eqref{eq:finite_dTY_identity} for the expenditure share, converts it into a quantity of public services at the wage and unit cost after automation, and reports the result for every scenario. Across the nine scenarios of Table~\ref{tab:scenario_grid}, the growth of public-service provision consistent with unchanged fiscal pressure ranges from 1.1 percent for a 1pp private-sector shock with no public automation to 19.9 percent for the symmetric 10pp shock. For details, see Table~\ref{tab:neutral_provision} in Appendix~\ref{app:neutral_provision}. In every case, services can grow, but by less than under demand-responsive provision.

As in the rest of our main analysis, these calculations measure how much additional provision automation can finance under the existing tax and transfer system. The calculation holds tax rates fixed and residual budget items such as cash transfers unchanged as shares of value added. Section~\ref{sec:transfers_extension} relaxes the assumption on transfers.

\subsection{Cross-country results}\label{sec:cross_country}

\begin{figure}[tbp]
\centering
\includegraphics[width=\textwidth]{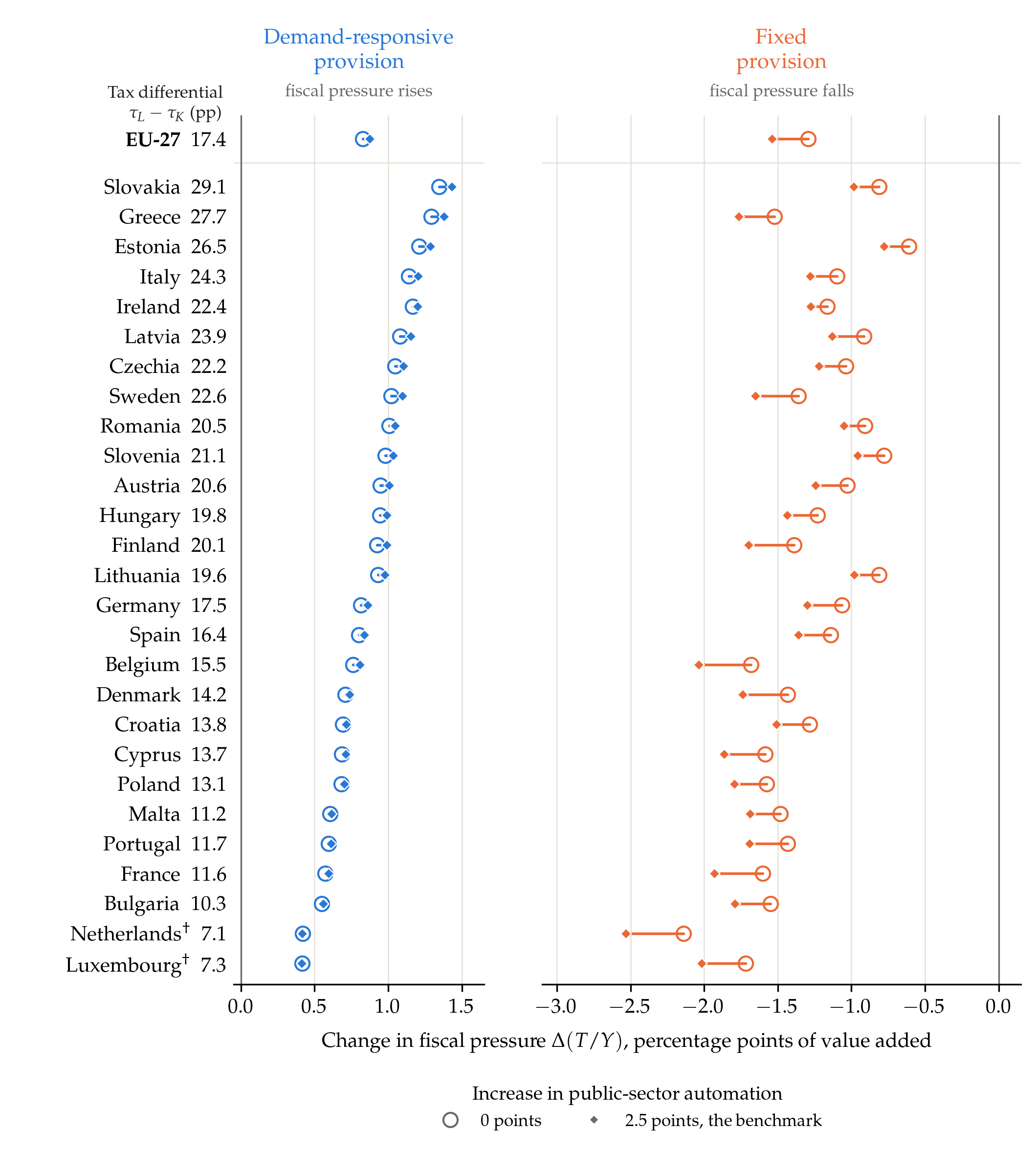}
\caption{Change in fiscal pressure due to a 5pp rise in the private-sector automation threshold}
\label{fig:cross_country_bar}
\begin{flushleft}
\scriptsize
Notes: Exact finite changes in percentage points of value added for a 5pp rise in the private-sector automation threshold. Both panels use the same horizontal scale. Open circles show no increase in public-sector automation and filled diamonds the benchmark increase of 2.5pp. The number after each country is its labor--capital tax differential $\tau_L-\tau_K$ in percentage points. The EU-27 aggregate is shown first. Member states follow in the order of Table~\ref{tab:cross_country_results}, ranked by the demand-responsive increase in the benchmark scenario. A dagger marks the Netherlands and Luxembourg, where the public-sector part of the shock lowers the rise in fiscal pressure by 0.004pp and 0.005pp. Parameters as in Tables~\ref{tab:eu_average} and~\ref{tab:cross_country_calibration}, with $A_K$ set for a 20 percent initial cost saving.
\end{flushleft}
\end{figure}

Figure~\ref{fig:cross_country_bar} shows the change in fiscal pressure for each member state, using the benchmark scenario with and without the public-sector part of the shock. In every country, fiscal pressure rises under demand-responsive provision and falls under fixed provision, under both assumptions about public-sector automation.\footnote{We apply the benchmark shock to each country's initial sector shares, labor shares and tax rates. Table~\ref{tab:cross_country_results} in Appendix~\ref{app:countries} reports the two contributions and the labor--capital tax differential for each country.} In the benchmark scenario, the demand-responsive increase ranges from 0.41pp in Luxembourg to 1.43pp in Slovakia, so the largest impact is roughly three and a half times the smallest.\footnote{The responses are only approximately proportional to the size of the shock. Doubling the demand-responsive benchmark response, for example, understates the exact response to a shock twice as large, the 10/5 row of Table~\ref{tab:scenario_grid}, by about 1 percent for the EU-27 and 4 percent for Ireland.}

\paragraph{The role of the tax differential.}
Most of the cross-country variation comes from the tax-base channel (see Table~\ref{tab:cross_country_results} in Appendix~\ref{app:countries} for detailed results). Its contribution ranges from 0.32pp in the Netherlands to 1.33pp in Slovakia. Expenditure contributions are smaller and vary less, between 0.07pp and 0.15pp, and show little relation to the labor--capital tax differential. The tax-base contribution $-(\tau_L-\tau_K)\Delta s_L$ is approximately proportional to the tax differential, because we apply the same shock to every country and labor shares fall by similar amounts. Slovakia, Greece, Estonia and Italy, with differentials of 24pp to 29pp, have some of the largest fiscal-pressure increases, while the Netherlands and Luxembourg, with differentials of about 7pp, have the smallest.
The ranking thus follows the tax differential closely, while it is only weakly related to the size of the public-service sector (see Table~\ref{tab:cross_country_calibration} in Appendix~\ref{app:countries}).

Ireland is the clearest deviation. It ranks fifth despite having the seventh-largest tax differential. One reason is that its small public-service sector gives the private-sector shock more weight, which makes direct displacement larger. Ireland's tax-base contribution still falls short of Latvia's, but its expenditure contribution is the largest in the sample because of its high initial private-sector automation threshold. With few private-sector tasks still performed by labor, the same 5pp shock automates a larger fraction of them, so wages and the unit cost of public services rise by more.\footnote{The small public-service share and the high threshold both partly reflect multinational profit shifting, which inflates measured value added in Ireland \citep{torslov2023missing}.}

\paragraph{Provision rules and public-sector automation.}
Figure~\ref{fig:cross_country_bar} also shows that, within each country, the provision rule matters far more than the sector split of the shock. The responses with and without public-sector automation nearly coincide under demand-responsive provision, whereas public-sector automation visibly deepens the fall under fixed provision. 

In the benchmark calibration, the public-sector part of the shock raises demand-responsive fiscal pressure in all countries except the Netherlands and Luxembourg, where it lowers the increase by 0.004pp and 0.005pp. This is because their small tax differentials let expenditure relief outweigh the tax-base effect of the additional displacement. The exception depends on the calibration, however.\footnote{With a 30 percent initial saving, which raises the expenditure relief from public-sector automation but not the displacement it adds, the public-sector part of the shock reduces the demand-responsive increase in seven countries. With the taxes on the stock of capital excluded from $\tau_K$, it adds to the increase in every country.}

Under fixed provision, the Netherlands, Belgium and Luxembourg have the largest fiscal-pressure reductions in the benchmark scenario. In the Netherlands and Luxembourg, small tax differentials limit the revenue loss that offsets the fall in the public-service share. In Belgium, a large public-service share generates a strong expenditure response. Estonia, Slovenia, Lithuania and Slovakia have the smallest reductions. Countries with large increases under demand-responsive provision thus tend to have small reductions under fixed provision.

A large tax differential therefore works against public finances under both provision rules. Despite its large tax differential, Greece has a relatively large fixed-provision improvement, because its high private-sector automation threshold, which partly reflects the mixed income of the self-employed, generates strong output growth.

\paragraph{Sensitivity of the country ranking.}
We check the stability of the ranking in the demand-responsive case under four alternative calibrations (for details see Appendix~\ref{app:countries}). These (i) exclude taxes on the stock of capital from $\tau_K$, (ii) use the country estimates of $\varepsilon$ in \citet{francois2023intratemporal} where available, (iii) treat payroll taxes as labor income, and (iv) treat two thirds of mixed income as labor income. The first three leave the ranking largely unchanged, with rank correlations of 0.90, 0.93 and 0.98 relative to the benchmark calibration, although Ireland, whose estimate of $\varepsilon$ exceeds one, falls from fifth to fifteenth under variant (ii). The fourth has a more meaningful impact on the rank order (the correlation is 0.76), and moves Greece from second to twenty-third. In all four cases, fiscal pressure still rises under demand-responsive provision and falls under fixed provision in every country.

\FloatBarrier
\section{Extensions}\label{sec:ext}

The results above rest on several simplifications. The government pays no cash transfers, all capital income accrues to domestic residents, firms rent capital at a fixed rental rate, and $\tau_K$ applies to the capital income generated in public-service production. Each factor is as productive in public services as in market production, tasks combine with unit elasticity, and AI replaces workers in existing tasks. Workers are of one type and firms earn no pure profits. This section examines departures from these assumptions in turn, beginning with those that matter most for the calibrated results.

\subsection{Transfers, pensions, and indexation}\label{sec:transfers_extension}

The model focuses on public services because they are the direct source of Baumol cost pressure. But modern welfare states also spend heavily on cash transfers, such as pensions and other social insurance benefits. Let $G$ denote this spending, measured in units of the private good, and let $s_G=G/Y$. Fiscal pressure in \eqref{eq:TY} then includes the additional term $s_G$, and \eqref{eq:dTY} gains the term $ds_G=s_G(d\log G-d\log Y)$. If benefits grow in step with value added, $s_G$ stays constant, so the changes in fiscal pressure reported in Section~\ref{sec:quant} apply equally to the budget including transfers. We consider two indexation rules instead. If real transfers are fixed in units of the private good, which corresponds to indexation to consumer prices, then $d\log G=0$, and output growth lowers transfer spending as a share of value added. If transfers are indexed to wages, then $d\log G=d\log W$ and, since $s_L=W/Y$, $ds_G=(s_G/s_L)\,ds_L$. The version of \eqref{eq:dTY} with wage-indexed transfers becomes
\[
d(T/Y)=(1+\tau_c)ds_E-\left[(\tau_L-\tau_K)-\frac{s_G}{s_L}\right]ds_L.
\]
Wage indexation ties the transfer share to the labor share, so in the cases above where $s_L$ falls, wage-indexed transfers fall relative to value added. If $s_G/s_L>\tau_L-\tau_K$, this transfer response is strong enough to reverse the sign of the term involving $ds_L$.

To gauge the size of the transfer response, suppose that all cash social benefits paid by general government in the EU-27, which amounted to 17.6 percent of value added in 2023, follow one of these rules.\footnote{These are social benefits other than social transfers in kind (item D.62), which were 15.9 percent of EU GDP in 2023 \citep{EurostatGov10aMain}. We convert this share to value added with the ratio of GDP to gross value added, 1.106 \citep{EurostatNama10Gdp}.} In the benchmark scenario, where fiscal pressure rises by 0.87pp under demand-responsive provision, $s_G$ then falls by about 2.0pp if benefits are fixed in real terms and 1.5pp if they follow wages. The wage-indexed case satisfies $s_G/s_L>\tau_L-\tau_K$ by a wide margin, since the ratio is roughly twice the labor--capital tax differential. More generally, in every scenario of Table~\ref{tab:scenario_grid} and under both indexation rules, the fall in $s_G$ is larger than the demand-responsive increase. In the benchmark scenario, benefits under these rules grow by at most 2.8 percent, while value added in units of the private good grows by 12.6 percent. Actual spending also depends on the number of recipients and on how new pensions are set. Finally, the calculations measure benefits before income tax, so accounting for these taxes would reduce the offset to fiscal pressure if effective tax rates on benefits stay unchanged.

\subsection{Foreign ownership of capital}\label{sec:foreign_ownership}

In a small open economy, part of the capital stock, including the capital that automation adds, may be foreign owned. This case is relevant for Europe if the AI systems that take over tasks are owned by firms outside the EU. Let $\omega\in[0,1]$ denote the domestic ownership share of capital income. The tax $\tau_K$ applies to residents' capital income, so capital-tax revenue is $\omega\tau_K RK$. We assume that part of private output is exported to pay foreign capital owners and that domestic residents receive no income from assets abroad. The household and government budgets then give
\[
C_m+P_sC_s=Y-(1-\omega)RK.
\]
Using $RK/Y=1-s_L$, fiscal pressure becomes
\[
\frac{T}{Y}=(1+\tau_c)s_E-\omega(\tau_K+\tau_c)-\bigl[\tau_L-\omega\tau_K+(1-\omega)\tau_c\bigr]s_L.
\]
Holding the public-service expenditure share fixed, a lower domestic ownership share makes fiscal pressure rise by more for a given fall in the labor share, provided capital income or consumption is taxed. A change in $\omega$ leaves wages and public-service unit costs unchanged at given automation thresholds and values of $R$ and $A_K$. Under demand-responsive provision, however, \eqref{eq:odds} must be combined with the resource identity above to determine $s_E$. Differentiating the expression for fiscal pressure, with $\omega$ now allowed to vary, gives
\[
d(T/Y)=(1+\tau_c)ds_E-\bigl[\tau_L-\omega\tau_K+(1-\omega)\tau_c\bigr]ds_L-(\tau_K+\tau_c)(1-s_L)d\omega.
\]
The last term captures the loss of capital-tax and consumption-tax revenue as ownership shifts abroad and is nonnegative when $d\omega<0$. Setting $\omega=1$ and $d\omega=0$ recovers \eqref{eq:dTY}.

Suppose residents keep their initial capital stock $K_0$ and foreign investors finance all of the capital that automation adds, with the gross rental rate $R$ unchanged. Then $\omega=K_0/K$ falls as $K$ rises, to about 0.8 in the EU-27 benchmark scenario of Section~\ref{sec:quant}. With $\tau_K$ levied on residents' capital income, as above, fiscal pressure rises by 2.9pp of value added under both provision rules. In the baseline model (Table~\ref{tab:scenario_grid}), by comparison, it rises by 0.87pp under demand-responsive provision and falls by 1.54pp under fixed provision. The two rules give similar results because the income from the new capital goes abroad, so residents' income grows only through wages. Demand-responsive provision then implies almost no growth in public services, and the expenditure share falls by about as much as under fixed provision. As before, the 2.9pp is a change in the overall deficit ratio only if other budget items remain constant as shares of domestic value added. This assumption matters more here, because domestic value added grows much faster than residents' income. If $\tau_K$ is instead levied where the capital is used, so that foreign owners pay it and receive their income net of tax, fiscal pressure rises by 0.8pp under demand-responsive provision and 0.2pp under fixed provision.\footnote{With the gross rental rate unchanged, this tax lowers foreign owners' net return. We do not model how their required return would affect capital costs and wages.} The 0.8pp rise is slightly below the increase in the baseline model, because the lower demand for public services more than offsets the consumption tax lost on the income paid abroad. Under fixed provision, the expenditure share and the labor share are the same as in the baseline model, so the whole difference in fiscal pressure is the consumption tax lost on that income.

\subsection{Capital supply}\label{sec:capital_supply}

In the baseline model, firms rent as much capital as they want at the world rental rate $R$. Appendix~\ref{app:fixed_capital} reports the opposite case for the benchmark scenario. The capital stock stays at its initial level, and the rental rate rises until firms' demand equals that stock. The fixed capital stock can be read as the short run and the baseline model as the long run, although the model does not describe the transition between them.

With the capital stock fixed, the rental rate increases from 0.158 to about 0.17, and the cost saving from automating a task shrinks from 20 percent to 3 or 4 percent. The wage drops by 9 to 10 percent rather than rising by 2.8 percent, because displaced workers compete for the remaining labor tasks while capital has become more expensive. Public services become cheaper as wages fall, so the expenditure contribution turns negative under demand-responsive provision. Under fixed provision it stays negative but is about half as large as in the baseline model, because value added in units of the private good no longer grows. Direct displacement is unchanged and the tax-base contribution hardly changes. Fiscal pressure rises by 0.47pp of value added under demand-responsive provision, compared with 0.87 in the baseline model, and falls by 0.34pp under fixed provision, compared with 1.54 (Table~\ref{tab:fixed_capital}), so both signs survive while the magnitudes shrink. The same calculation works for the six scenarios of Table~\ref{tab:scenario_grid} in which the private-sector threshold rises by 1pp or 5pp, whereas the three 10pp scenarios have no solution in which capital stays cheaper than labor in automatable tasks ($R/A_K<1$).

\subsection{Taxable public-sector capital}

The remaining extensions retain the fixed rental rate of Sections~\ref{sec:model} to~\ref{sec:cal}. The baseline model applies $\tau_K$ to the entire capital share of value added, including capital used in public-service production. However, under the European System of Accounts (ESA 2010), non-market government output is valued at cost and includes no net return on government-owned capital. The capital component of its value added therefore consists mainly of consumption of fixed capital on these assets, which is not taxable private capital income. As an extreme case, suppose that none of the capital income of public-service production is taxed. Capital-tax revenue as a share of value added is then $\tau_KI_m(1-s_E)$ in place of $\tau_K(1-s_L)$, so \eqref{eq:dTY} gains the term $\tau_K\,d(I_ss_E)$. At an unchanged capital-tax rate, this term raises fiscal pressure whenever $I_ss_E$, the capital income of public-service production as a share of value added, increases. Preserving initial capital-tax receipts, however, raises $\tau_K$ from 0.196 to about 0.214, and we hold this rate fixed during the automation shock. The higher rate narrows the labor--capital tax differential and so reduces the tax-base contribution of a fall in the labor share. In the benchmark scenario, fiscal pressure rises by 0.89pp of value added under demand-responsive provision and falls by 1.65pp under fixed provision, compared with 0.87 and 1.54 in the baseline model. Under demand-responsive provision, public-service capital income rises as a share of value added, and excluding it from taxation raises fiscal pressure slightly more than the higher capital-tax rate lowers it. Under fixed provision, public-service capital income falls as a share of value added, so both changes make fiscal pressure fall further.

\subsection{Production assumptions}\label{sec:robustness}

In the baseline model, each factor is as productive in public services as in market production. The part of Appendix~\ref{app:factor_productivity} on automation relaxes this assumption and lets task productivity differ across sectors. If capital remains cheaper in automatable tasks in both sectors, the labor-share identity and the fiscal-pressure decomposition are unchanged. Holding productivities fixed, automation still raises the wage, and it also raises the public-service unit cost if capital's productivity relative to labor is no greater in public services than in market production.

The baseline model also assumes Cobb-Douglas task aggregation, with $\sigma=1$, and Appendix~\ref{app:CES} extends it to CES task aggregation. Whenever a positive, finite equilibrium exists ($I_m(R/A_K)^{1-\sigma}<1$), automation continues to raise the wage and the unit cost of public services, and the fiscal-pressure decomposition retains the same structure. However, when $\sigma<1$, automation need not lower the aggregate labor share at a fixed expenditure share, so the fiscal sign results do not carry over automatically. At the EU-27 values of Section~\ref{sec:cal}, with the initial thresholds, expenditure share and effective cost of capital held fixed, the displacement term is nevertheless positive for every $\sigma$ below one (Appendix~\ref{app:CES}), so Results~\ref{prop:fiscal_general} and~\ref{prop:demand} continue to apply there.

\subsection{Labor-augmenting AI and new tasks}\label{sec:augmentation}

Automation is not the only way in which AI can affect production. In the baseline model, machines take over tasks that workers used to do, but AI can instead make workers more productive in the tasks they keep. The final part of Appendix~\ref{app:factor_productivity}, on labor-augmenting AI, works this out with the automation thresholds fixed, so that no tasks move from workers to machines. Suppose public services follow demand and, as in our benchmark calibration, are gross complements to private consumption. The fiscal consequences then depend on where the productivity gain occurs. If private-sector workers become more productive, wages rise across the economy, and public services, which are produced mostly with labor, become more expensive. This is Baumol's cost disease in its classic form. Spending on public services rises as a share of value added, and so does fiscal pressure. However, the sectoral labor shares are unchanged, so there is no displacement effect on the tax base. If instead public-sector workers become more productive, for example because AI helps teachers prepare lessons or helps doctors reach a diagnosis, public services become cheaper to provide and fiscal pressure falls. Under fixed provision, either productivity gain lowers fiscal pressure, because output grows faster than the cost of the fixed service bundle. The contrast with automation is sharpest in the public sector, where augmentation lowers fiscal pressure under both provision rules, whereas public-sector automation that catches up with the private sector raises it under demand-responsive provision in the EU-27 benchmark calibration.

AI may also create new tasks that only workers can do. The model has a fixed set of tasks, so new tasks enter it only indirectly, through the net interpretation of automation changes in Section~\ref{sec:model}. New tasks would work against displacement, because by giving workers something else to do they can keep the labor share, and with it the labor-income tax base, from falling. Studying them would require a mechanism that adds tasks on the labor side, as in \citet{acemoglu2018race}, which our model does not have.

\subsection{Heterogeneous labor and AI rents}\label{sec:labor_rents}

With one labor type, all labor income enters the fiscal decomposition through the aggregate share $s_L$. The model therefore cannot show how automation affects different groups of workers. A richer model would distinguish skill groups, occupations or sectors of employment, which matters because AI may substitute for some tasks while complementing others. The tax-base channel would then depend not only on the aggregate labor share but also on how each component of labor income is taxed. Progressive taxation, ceilings on social contributions and sector-specific wage setting could make the revenue effect larger or smaller than $ds_L$ alone suggests. Even with one labor type, a progressive schedule whose thresholds do not rise with wages would raise the effective tax rate on labor as wages grow, which would work against the tax-base channel.

Pure profits are absent from the baseline model as well. We do not model markups by changing the pricing equations. Instead, suppose that some income appears as AI rents $\Pi$ and is taxed at rate $\tau_\Pi$. Let $s_\Pi=\Pi/Y$ be the profit share, so that normal capital income is now a share $1-s_L-s_\Pi$ of value added. As shares of value added, capital-tax revenue falls by $\tau_Ks_\Pi$ and profit-tax revenue rises by $\tau_\Pi s_\Pi$. Fiscal pressure therefore changes by $-(\tau_\Pi-\tau_K)s_\Pi$. With fixed tax rates, \eqref{eq:dTY} becomes
\[
d(T/Y)=(1+\tau_c)ds_E-(\tau_L-\tau_K)ds_L-(\tau_\Pi-\tau_K)ds_\Pi,
\]
where we take $s_\Pi$ and $ds_\Pi$ as given. If AI raises the profit share, this lowers fiscal pressure when profits are taxed more heavily than normal capital income and raises it when they are taxed less heavily.

\section{Conclusion}\label{sec:concl}

This paper asks whether AI may ease fiscal pressure in countries that tax labor more heavily than capital and fund public services from taxes. In our benchmark calibration, automation shifts income toward more lightly taxed capital while raising value added by 12 percent under demand-responsive provision. In this case the deficit rises by nearly 1pp of value added, mostly because of the tax-base channel. If public services were held fixed in real terms, the deficit would instead fall by 1.5pp of value added in the EU-27 benchmark scenario, because the fall in the public-service share outweighs the fall in tax receipts relative to value added. This would, however, imply a falling expenditure share for public services even though incomes grow and the relative cost of public services rises. Moreover, if foreign investors own all the capital that automation adds, the income paid abroad escapes the consumption tax and (under residence-based taxation) the capital tax, so fiscal pressure rises even under fixed provision.

We do not model how governments should respond, but our results highlight that increased automation is likely to require a fall in the public-service expenditure share, higher tax revenue as a share of value added, or user fees. Importantly, growth by itself does not remove this need. When public services follow demand, public-service spending rises with income, so its share of value added does not fall as the economy grows. Automating public services faster might seem a natural way out, because it holds down their cost. Yet, as shown in Sections~\ref{sec:quant} and~\ref{sec:cross_country}, it also lowers the labor share and shifts income toward capital. In the benchmark scenario, this tax-base effect outweighs the expenditure saving under demand-responsive provision, so additional public-sector automation raises fiscal pressure in the EU-27 and in all but two member states.

To keep the model tractable, we make several simplifying assumptions, relying on a representative household with one labor type and full employment. Future work could relax these assumptions, and our model provides a reasonable starting point for such extensions. Most notably, it seems plausible that automation may reduce employment, which would lower labor-tax receipts and increase benefit spending. In this respect, our results are likely to understate the fiscal pressure arising from automation. We also take automation and provision rules as given and do not solve for optimal policy. An interesting extension would be to let taxes affect the pace of automation.
\FloatBarrier
\setlength{\bibsep}{7pt plus 1pt}\bibliographystyle{econ}
\bibliography{refs.bib}

\newpage 
\appendix

\section{Model derivations}

\subsection{Task cost minimization}\label{app:task_equilibrium}

This appendix derives the unit cost in \eqref{eq:price_j}, the task expenditure condition in \eqref{eq:per_task_expenditure}, and the factor demand formulas in \eqref{eq:factor_demands} from the Cobb-Douglas task aggregator \eqref{eq:aggregator}. For a given sector $j$ and target output $Y_j$, the producer solves
\begin{equation}\label{eq:task_cost_min_app}
\min_{\{y_j(i)>0\}}\int_0^1 p_j(i)y_j(i)\,di
\quad\text{s.t.}\quad
\int_0^1\log y_j(i)\,di\geq\log Y_j.
\end{equation}
Because task prices are positive, the constraint binds at the optimum. Let $\lambda_j>0$ denote the multiplier. For each task $i$, the first-order condition is $p_j(i)=\lambda_j/y_j(i)$, or
\begin{equation}\label{eq:task_expenditure_app}
p_j(i)y_j(i)=\lambda_j
\qquad\text{for all }i\in[0,1].
\end{equation}
Integrating $\log y_j(i)=\log\lambda_j-\log p_j(i)$ over $i\in[0,1]$ and using the constraint in \eqref{eq:task_cost_min_app} gives
\[
\log\lambda_j=\log Y_j+\int_0^1\log p_j(i)\,di.
\]
Thus $\lambda_j=P_jY_j$, where
\[
P_j\equiv\exp\!\int_0^1 \log p_j(i)\,di
\]
is the sectoral unit cost. Since the task interval has measure one, minimized total task expenditure is
\[
\int_0^1 p_j(i)y_j(i)\,di=\int_0^1\lambda_j\,di=\lambda_j=P_jY_j.
\]
Producing one unit of sectoral output therefore costs $P_j$, and \eqref{eq:task_expenditure_app} is equivalent to $p_j(i)y_j(i)=P_jY_j$, so cost minimization equalizes expenditure across tasks.

By \eqref{eq:task_prices}, $p_j(i)=R/A_K$ on $[0,I_j)$ and $p_j(i)=W$ on $[I_j,1]$, so $\int_0^1\log p_j(i)\,di=I_j\log(R/A_K)+(1-I_j)\log W$ and $P_j=(R/A_K)^{I_j}W^{1-I_j}$, as in \eqref{eq:price_j}. Combining these task prices with \eqref{eq:per_task_expenditure} gives $y_j(i)=A_KP_jY_j/R$ for tasks performed by capital and $y_j(i)=P_jY_j/W$ for tasks performed by labor. With $R/A_K<W$, firms use only capital below the threshold and only labor above it, so $k_j(i)=y_j(i)/A_K$ for $i<I_j$ and $l_j(i)=y_j(i)$ for $i\geq I_j$. Integrating over the two task intervals gives the factor demand equations in \eqref{eq:factor_demands}.

\subsection{The expenditure share under fixed provision}\label{app:fixed_share}

Under fixed public-service provision, $C_s=\widehat{C}_s$. The expenditure share \eqref{eq:econ_shares} is
\[
s_E=\frac{P_s\widehat{C}_s}{Y_m+P_s\widehat{C}_s}
=
\frac{P_s\widehat{C}_s/Y_m}{1+P_s\widehat{C}_s/Y_m}.
\]
Since $x/(1+x)$ is increasing for $x>0$, it suffices to show $d(P_s\widehat{C}_s/Y_m)<0$. Equation~\eqref{eq:factor_demands} gives $L_j=(1-I_j)P_jY_j/W$. Since $P_m=1$ and $Y_s=\widehat{C}_s$, it follows that $Y_m=WL_m/(1-I_m)$ and $P_s\widehat{C}_s=WL_s/(1-I_s)$. The latter can be written as $L_s=(1-I_s)\psi$, where
\[
\psi\equiv\frac{P_s\widehat{C}_s}{W}=\widehat{C}_s\left(\frac{R}{A_KW}\right)^{I_s}.
\]
The second equality uses $P_s=(R/A_K)^{I_s}W^{1-I_s}$ from \eqref{eq:price_j} with $j=s$. Since $L_m=1-L_s=1-\psi(1-I_s)$,
\begin{equation}\label{eq:sE_ratio}
\frac{P_s\widehat{C}_s}{Y_m}=\frac{\psi(1-I_m)}{1-\psi(1-I_s)},
\end{equation}
with positive numerator $\psi(1-I_m)$ and positive denominator $L_m$.

To sign $d(P_s\widehat{C}_s/Y_m)$, first take logs of $\psi$. With $q\equiv\log(A_KW/R)>0$ by \eqref{eq:wage} and Assumption~\ref{ass:initial}, $\log\psi=\mathrm{const}-I_s q$, so
\[
d\log\psi=-q\,dI_s-I_s\,d\log W=-q\,dI_s-\frac{I_s q}{1-I_m}\,dI_m<0,
\]
where the second equality uses \eqref{eq:dlogW} and the inequality uses $q>0$ and $I_s>0$. It holds for any $dI_m\geq0$ and $dI_s\geq0$ that are not both zero, so the argument below also covers changes in which public-sector automation advances by more than private-sector automation. Hence $d\psi<0$. Differentiating the numerator and denominator of \eqref{eq:sE_ratio},
\[
\begin{aligned}
d[\psi(1-I_m)]&=(1-I_m)\,d\psi-\psi\,dI_m<0,\\
d[1-\psi(1-I_s)]&=-(1-I_s)\,d\psi+\psi\,dI_s>0.
\end{aligned}
\]
For $a>0$ and $b>0$, the ratio $a/b$ is increasing in $a$ and decreasing in $b$. In \eqref{eq:sE_ratio}, the numerator decreases and the denominator increases. Hence
\[
d(P_s\widehat{C}_s/Y_m)<0.
\]
Therefore $ds_E<0$, so by \eqref{eq:dsE_fixed} output grows faster than the unit cost of public services.

\subsection{CES preferences and demand-responsive provision}\label{app:ces_eta}

This appendix specifies household demand under CES preferences for private consumption and public services, at a fixed consumption-tax rate $\tau_c$. The parameter $\varepsilon$ is the elasticity of substitution between the two goods.

Let preferences over private consumption $C_m$ and public-service consumption $C_s$ be
\begin{equation}\label{eq:utility}
u(C_m,C_s)=
\left[
\alpha^{1/\varepsilon}C_m^{(\varepsilon-1)/\varepsilon}
+(1-\alpha)^{1/\varepsilon}C_s^{(\varepsilon-1)/\varepsilon}
\right]^{\varepsilon/(\varepsilon-1)},
\end{equation}
with $\alpha\in(0,1)$ the weight on private consumption and $\varepsilon>0$, $\varepsilon\neq1$. The Cobb-Douglas case is given by
\[
\lim_{\varepsilon\to1}u(C_m,C_s)
=
\alpha^{-\alpha}(1-\alpha)^{-(1-\alpha)}C_m^\alpha C_s^{1-\alpha},
\]
which differs from the usual Cobb-Douglas utility only by multiplication by a positive constant. Under the demand-responsive rule, $C_s$ is set so that the ratio $C_s/C_m$ equals the one households would choose if public services cost $P_s$ per unit and private consumption cost $1+\tau_c$. Because these preferences are homothetic, this ratio does not depend on household income. The household's first-order condition equates the marginal rate of substitution to the relative price,
\[
\frac{\partial u/\partial C_s}{\partial u/\partial C_m}
=\left(\frac{1-\alpha}{\alpha}\right)^{1/\varepsilon}\!\left(\frac{C_s}{C_m}\right)^{-1/\varepsilon}
=\frac{P_s}{1+\tau_c},
\]
which gives
\begin{equation}\label{eq:demand_responsive}
\frac{C_s}{C_m}=\frac{1-\alpha}{\alpha}\,(1+\tau_c)^\varepsilon P_s^{-\varepsilon}.
\end{equation}
When households face the same price for both goods, $P_s=1+\tau_c$, \eqref{eq:demand_responsive} gives $C_s/C_m=(1-\alpha)/\alpha$. In this hypothetical choice, households would spend share $\alpha$ on private consumption, including consumption tax, and share $1-\alpha$ on public services.
Using $s_E/(1-s_E)=P_sC_s/C_m$ (which follows from $Y=C_m+P_sC_s$ in \eqref{eq:gdp_resource} and $s_E=P_sC_s/Y$), the ratio of public to private expenditure before consumption taxes satisfies
\begin{equation}\label{eq:demand_share}
\frac{s_E}{1-s_E}=\frac{1-\alpha}{\alpha}\,(1+\tau_c)^\varepsilon P_s^{1-\varepsilon}.
\end{equation}
This is the provision schedule \eqref{eq:odds} with $\Omega(P_s)=\frac{1-\alpha}{\alpha}\,(1+\tau_c)^\varepsilon P_s^{1-\varepsilon}$. At fixed $\tau_c$, log-differentiating \eqref{eq:demand_share} gives
\begin{equation}\label{eq:dlog_share}
d\log\frac{s_E}{1-s_E}=(1-\varepsilon)\,d\log P_s,
\end{equation}
so, using the definition in \eqref{eq:etaP},
\[
\eta_P
=
\frac{d\log\!\left(s_E/(1-s_E)\right)}{d\log P_s}
=
1-\varepsilon.
\]
At the Cobb-Douglas limit $\varepsilon=1$, \eqref{eq:demand_share} becomes $s_E/(1-s_E)=[(1-\alpha)/\alpha](1+\tau_c)$, so the expenditure ratio does not respond to $P_s$ and $\eta_P=0$.

\subsection{Provision that leaves fiscal pressure unchanged}\label{app:neutral_provision}

Let $I_m'$ and $I_s'$ be the automation thresholds after a finite change, $g'=I_m'-I_s'$, and $W'$ and $P_s'$ the wage and unit cost from \eqref{eq:wage} and \eqref{eq:Ps} at the new thresholds. For any expenditure share $s_E'$ after the change, the labor-share identity gives $s_L'=1-I_m'+g's_E'$, and \eqref{eq:finite_dTY_identity} gives the change in fiscal pressure. Setting $\Delta(T/Y)=0$ and solving for the expenditure share,
\[
\tilde{s}_E=\frac{(1+\tau_c)s_E+(\tau_L-\tau_K)(1-I_m'-s_L)}{(1+\tau_c)-(\tau_L-\tau_K)g'},
\]
where $s_E$ and $s_L$ are the initial shares. The denominator is positive whenever $(\tau_L-\tau_K)g'<1+\tau_c$, which holds for every scenario we consider, so the solution is unique. The corresponding quantity of public services is $\tilde{C}_s=\tilde{s}_EY'/P_s'$, with $Y'=W'/s_L'$ at $L=1$, and $\tilde{C}_s/C_s-1$ is the growth of provision that leaves fiscal pressure unchanged. By \eqref{eq:finite_dTY_decomp}, fiscal pressure increases with $s_E'$ at given thresholds, with slope $1+\tau_c-(\tau_L-\tau_K)g'>0$. Provision below $\tilde{C}_s$ therefore lowers it, while provision above it raises it. Table~\ref{tab:neutral_provision} reports $\tilde{s}_E$ and $\tilde{C}_s/C_s-1$ for the nine scenarios of Table~\ref{tab:scenario_grid}, next to the growth of public services under demand-responsive provision, and Section~\ref{sec:quant} discusses them.

\begin{table}[htbp]
\centering
\caption{Growth of public services that leaves fiscal pressure unchanged, EU-27, 2023}
\label{tab:neutral_provision}
\footnotesize
\begin{tabular}{@{}l rr r rr@{}}
\toprule
& & & & \multicolumn{2}{c}{Growth of public services (\%)} \\
\cmidrule(lr){5-6}
Scenario & $\Delta I_m$ & $\Delta I_s$ & \begin{tabular}[b]{@{}c@{}}Expenditure share\\$\tilde{s}_E$ (\% of value added)\end{tabular} & \begin{tabular}[b]{@{}c@{}}Unchanged\\fiscal pressure\end{tabular} & \begin{tabular}[b]{@{}c@{}}Demand-\\responsive\end{tabular} \\
\midrule
Private 1, Public 0 & 1 & 0 & 18.2 & 1.1 & 1.8 \\
Private 1, Public 0.5 & 1 & 0.5 & 18.2 & 1.3 & 2.1 \\
Symmetric 1 & 1 & 1 & 18.1 & 1.6 & 2.4 \\
Private 5, Public 0 & 5 & 0 & 17.7 & 6.0 & 9.7 \\
Private 5, Public 2.5 & 5 & 2.5 & 17.6 & 7.3 & 11.3 \\
Symmetric 5 & 5 & 5 & 17.5 & 8.6 & 13.0 \\
Private 10, Public 0 & 10 & 0 & 17.0 & 13.4 & 21.6 \\
Private 10, Public 5 & 10 & 5 & 16.9 & 16.6 & 25.6 \\
Symmetric 10 & 10 & 10 & 16.8 & 19.9 & 29.8 \\
\bottomrule
\end{tabular}
\begin{flushleft}
\scriptsize
Notes: $\Delta I_m$ and $\Delta I_s$ in percentage points. $\tilde{s}_E$ is the public-service expenditure share, in percent of value added, at which the change in fiscal pressure is zero after the shock, from the formula above. The initial share is 18.3 percent. The next column is the percentage growth in the real quantity of public services that this share finances, $\tilde{C}_s/C_s-1$, and the last column the growth under demand-responsive provision, from the same equilibria as Table~\ref{tab:scenario_grid}. Under fixed provision the quantity does not change, so its growth is zero in every scenario. Tax rates are held fixed and the budget items outside the model are held constant as shares of value added. Parameters as in Table~\ref{tab:eu_average}, with $A_K$ set for a 20 percent initial cost saving.
\end{flushleft}
\end{table}

\FloatBarrier
\section{Alternative production assumptions}

\subsection{Factor and sector differences in task productivity}\label{app:factor_productivity}

In the main text, capital has the same productivity $A_K$ in both sectors and labor productivity is one everywhere. This appendix lets both differ across sectors and asks what changes. Three results follow. First, the labor-share identity and the fiscal-pressure decomposition are the same as in the main text, as long as capital is still the cheaper way to perform automatable tasks in both sectors. Second, for the automation changes studied in Section~\ref{sec:cs}, wages still rise, and the cost of public services still rises provided capital's productivity advantage is no larger in public services than in market production. Third, the appendix works out what happens when AI makes workers more productive instead of replacing them, which is the case used in Section~\ref{sec:augmentation}.

\paragraph{Setup.}
To generalize the task technology in \eqref{eq:task_prod}, let $A_{Kj},A_{Lj}>0$ and
\[
y_j(i)=
\begin{cases}
A_{Kj}k_j(i)+A_{Lj}l_j(i) & \text{if } i<I_j,\\
A_{Lj}l_j(i) & \text{if } i\geq I_j,
\end{cases}
\qquad j\in\{m,s\}.
\]
Producing one unit of task output costs $R/A_{Kj}$ with capital and $W/A_{Lj}$ with labor. We consider the case in which capital is cheaper in both sectors, so firms use only capital below the threshold and only labor above it. Applying the same cost minimization argument as in Appendix~\ref{app:task_equilibrium}, the sectoral unit cost generalizes \eqref{eq:price_j} to
\begin{equation}\label{eq:factor_specific_price}
P_j
=
\left(\frac{R}{A_{Kj}}\right)^{I_j}
\left(\frac{W}{A_{Lj}}\right)^{1-I_j}.
\end{equation}
Capital is cheaper in both sectors when
\begin{equation}\label{eq:factor_specific_cost_saving}
\frac{R}{A_{Kj}}<\frac{W}{A_{Lj}},
\qquad j\in\{m,s\}.
\end{equation}

With Cobb-Douglas task aggregation, the cost minimization argument again equalizes expenditure across tasks. Since tasks in $[0,I_j)$ are performed by capital and tasks in $[I_j,1]$ are performed by labor, the sectoral cost shares are still
\[
s_K^j=I_j,
\qquad
s_L^j=1-I_j.
\]
This is the same expression as \eqref{eq:cost_shares}.
As a result, the labor-share identity \eqref{eq:sL_identity} and the fiscal-pressure decomposition \eqref{eq:dTY_decomp} remain intact whenever capital is cheaper in both sectors. However, productivity differences affect the responses of $W$ and $P_s$.

Setting $j=m$ in \eqref{eq:factor_specific_price} and imposing the numeraire normalization used in \eqref{eq:wage}, $P_m=1$, implies
\begin{equation}\label{eq:factor_specific_wage}
W
=
A_{Lm}\left(\frac{A_{Km}}{R}\right)^{I_m/(1-I_m)}.
\end{equation}
Substituting the wage in \eqref{eq:factor_specific_wage} into \eqref{eq:factor_specific_cost_saving} gives $R<A_{Km}$ for the market sector and $R/A_{Ks}<(A_{Lm}/A_{Ls})(A_{Km}/R)^{I_m/(1-I_m)}$ for the public sector. When productivities are the same across sectors, both conditions reduce to $R<A_{Km}$.
Differentiating \eqref{eq:factor_specific_wage} gives
\begin{equation}\label{eq:factor_specific_wage_diff}
d\log W
=
d\log A_{Lm}
+
\frac{I_m}{1-I_m}(d\log A_{Km}-d\log R)
+
\frac{\log(A_{Km}/R)}{(1-I_m)^2}\,dI_m.
\end{equation}
Setting $j=s$ in \eqref{eq:factor_specific_price} and differentiating gives
\begin{equation}\label{eq:factor_specific_Ps_diff}
\begin{split}
d\log P_s
&=
I_s(d\log R-d\log A_{Ks})
+(1-I_s)(d\log W-d\log A_{Ls})\\
&\quad+
\left[
\log\frac{R}{A_{Ks}}
-
\log\frac{W}{A_{Ls}}
\right]dI_s.
\end{split}
\end{equation}
Substituting \eqref{eq:factor_specific_wage_diff} into \eqref{eq:factor_specific_Ps_diff} gives the response of $P_s$.

\paragraph{Automation with sector-specific productivities.}
Consider the automation changes in Section~\ref{sec:cs}, holding all productivity parameters, the rental rate and tax rates fixed. Define
\[
\widetilde q_j\equiv
\log\frac{W/A_{Lj}}{R/A_{Kj}},
\qquad j\in\{m,s\}.
\]
Condition \eqref{eq:factor_specific_cost_saving} gives $\widetilde q_j>0$, and by \eqref{eq:factor_specific_wage}, $\widetilde q_m=\log(A_{Km}/R)/(1-I_m)$. Equations \eqref{eq:factor_specific_wage_diff} and \eqref{eq:factor_specific_Ps_diff} then imply
\[
d\log W=\frac{\widetilde q_m}{1-I_m}\,dI_m>0,
\qquad
d\log P_s=
\frac{1-I_s}{1-I_m}\,\widetilde q_m\,dI_m-\widetilde q_s\,dI_s.
\]
The positive wage response in Result~\ref{prop:prices} therefore survives. The public-service cost response remains positive if and only if
\[
\frac{1-I_s}{1-I_m}\,\widetilde q_m\,dI_m>\widetilde q_s\,dI_s.
\]
A sufficient condition is $\widetilde q_s\leq \widetilde q_m$, which is equivalent to
\[
\frac{A_{Ks}}{A_{Ls}}\leq \frac{A_{Km}}{A_{Lm}}.
\]
The productivity parameters determine the relative sizes of $\widetilde q_m$ and $\widetilde q_s$. The positive $P_s$ response can fail only if capital's productivity advantage in public services is large enough for the direct public-sector cost reduction to dominate the wage effect.

\paragraph{Labor-augmenting AI.}
To isolate the effects of higher labor productivity in either sector, hold $I_m$, $I_s$, $A_{Km}$, $A_{Ks}$, $R$ and tax rates fixed, and maintain \eqref{eq:factor_specific_cost_saving}. The sectoral labor shares are unchanged. Hence $D=0$ and $ds_L=g\,ds_E$, so in both cases \eqref{eq:dTY_decomp} reduces to
\[
d(T/Y)=\bigl[1+\tau_c-(\tau_L-\tau_K)g\bigr]ds_E.
\]
The coefficient on $ds_E$ is positive under the maintained assumptions.

An increase in private-sector labor productivity $A_{Lm}$, holding $A_{Ls}$ fixed, gives
\[
d\log W=d\log A_{Lm}>0,
\qquad
d\log P_s=(1-I_s)\,d\log A_{Lm}>0.
\]
An increase in public-sector labor productivity $A_{Ls}$, holding $A_{Lm}$ fixed, instead gives
\[
d\log W=0,
\qquad
d\log P_s=-(1-I_s)\,d\log A_{Ls}<0.
\]
Under demand-responsive provision, \eqref{eq:dsE_eta} implies that private-sector labor productivity growth raises fiscal pressure when $\eta_P>0$ and lowers it when $\eta_P<0$. Public-sector labor productivity growth has the opposite effect. Fiscal pressure is unchanged in either case when $\eta_P=0$.

Under fixed provision, combining \eqref{eq:dsE_fixed} with $Y=W/s_L$, $ds_L=g\,ds_E$ and \eqref{eq:sL_identity} gives
\[
ds_E=\frac{s_Es_L}{1-I_m}\bigl(d\log P_s-d\log W\bigr)<0.
\]
The inequality holds for either productivity increase because $d\log P_s<d\log W$. Thus fiscal pressure falls in both cases.

\subsection{CES task aggregation}\label{app:CES}

The main text uses Cobb-Douglas task aggregation, which makes the sectoral labor cost share in \eqref{eq:cost_shares} equal to $s_L^j=1-I_j$. To see which results depend on the Cobb-Douglas specification, suppose instead that task outputs are aggregated by
\[
Y_j=
\left[
\int_0^1 y_j(i)^{(\sigma-1)/\sigma}\,di
\right]^{\sigma/(\sigma-1)}.
\]
The parameter $\sigma>0$ is the elasticity of substitution across task outputs. The Cobb-Douglas aggregator in \eqref{eq:aggregator} is the limit as $\sigma\to 1$. Note that $\sigma$ is not an elasticity of substitution between capital and labor within a task, since the two remain perfect substitutes in automatable tasks, as in \eqref{eq:task_prod}. For $R/A_K<W$, cost minimization assigns tasks below $I_j$ to capital and the remaining tasks to labor.

Applying the same cost minimization argument as in Appendix~\ref{app:task_equilibrium}, now using the CES aggregator above, the sectoral unit cost generalizes \eqref{eq:price_j} to
\begin{equation}\label{eq:ces_price}
P_j
=
\left[I_j(R/A_K)^{1-\sigma}+(1-I_j)W^{1-\sigma}\right]^{1/(1-\sigma)}.
\end{equation}
The corresponding factor cost shares generalize \eqref{eq:cost_shares} to
\begin{equation}\label{eq:ces_shares}
s_K^j
=
\frac{I_j(R/A_K)^{1-\sigma}}{I_j(R/A_K)^{1-\sigma}+(1-I_j)W^{1-\sigma}},
\qquad
s_L^j
=
\frac{(1-I_j)W^{1-\sigma}}{I_j(R/A_K)^{1-\sigma}+(1-I_j)W^{1-\sigma}}.
\end{equation}
Unlike in the Cobb-Douglas case, the sectoral labor cost share $s_L^j$ is not simply $1-I_j$. It depends not only on the automation threshold $I_j$, but also on the wage $W$ and the capital cost $R/A_K$.
To simplify notation, write the bracketed term in \eqref{eq:ces_price} as $B_j$ and denote the CES labor cost share by $\ell_j$:
\[
B_j\equiv I_j(R/A_K)^{1-\sigma}+(1-I_j)W^{1-\sigma},
\qquad
\ell_j\equiv s_L^j
=
\frac{(1-I_j)W^{1-\sigma}}{B_j}.
\]
As in \eqref{eq:wage}, normalize the private-good price to $P_m=1$. Then $B_m=1$, so
\begin{equation}\label{eq:ces_wage_norm}
I_m(R/A_K)^{1-\sigma}+(1-I_m)W^{1-\sigma}=1.
\end{equation}
Equation~\eqref{eq:ces_wage_norm} plays the same role as \eqref{eq:wage} and determines $W$ given $R$, $A_K$ and $I_m$. Solving for $W^{1-\sigma}$ gives
\[
W^{1-\sigma}=\frac{1-I_m(R/A_K)^{1-\sigma}}{1-I_m}.
\]
For a positive solution $W$ to exist, we need $1-I_m(R/A_K)^{1-\sigma}>0$, or $I_m(R/A_K)^{1-\sigma}<1$. We assume this restriction holds.
The market-sector unit cost in \eqref{eq:ces_price} lies strictly between $R/A_K$ and $W$ when they differ. Thus, with $R/A_K<1$ and $P_m=1$, the solution satisfies $W>1>R/A_K$. As in footnote~\ref{fn:capital_cheaper}, $W\leq R/A_K$ would make every task's unit cost equal to $W$ and imply $P_m=W=1>R/A_K$, a contradiction.

Differentiating \eqref{eq:ces_wage_norm} gives
\begin{equation}\label{eq:ces_wage_response}
d\log W
=
\frac{1-\left(R/(A_KW)\right)^{1-\sigma}}{(1-I_m)(1-\sigma)}\,dI_m.
\end{equation}
As $\sigma\to 1$, \eqref{eq:ces_wage_response} converges to the Cobb-Douglas wage response in \eqref{eq:dlogW}. For $\sigma\neq1$ and $dI_m>0$, the wage response in \eqref{eq:ces_wage_response} is positive if and only if
\[
\frac{1-\left(R/(A_KW)\right)^{1-\sigma}}{1-\sigma}>0.
\]
Since $W>R/A_K$, this condition holds.

The response of the public-service unit cost $P_s$, corresponding to \eqref{eq:dlogPs}, is
\begin{equation}\label{eq:ces_price_response}
d\log P_s
=
\ell_s\,d\log W
+
\frac{(R/A_K)^{1-\sigma}-W^{1-\sigma}}{(1-\sigma)B_s}\,dI_s.
\end{equation}
As $\sigma\to1$, \eqref{eq:ces_price_response} converges to $d\log P_s=(1-I_s)\,d\log W-\log(A_KW/R)\,dI_s$, which is equivalent to \eqref{eq:dlogPs}. In both specifications, higher wages raise the cost of public services while public-sector automation lowers it. Substituting \eqref{eq:ces_wage_response} into \eqref{eq:ces_price_response} gives
\[
d\log P_s
=
\frac{W^{1-\sigma}-(R/A_K)^{1-\sigma}}{(1-\sigma)B_s}\,
\frac{1}{1-I_m}
\left[g\,dI_s+(1-I_s)\,dg\right].
\]
The first factor is positive whenever $W>R/A_K$, and the bracket is the same as in \eqref{eq:dlogPs}. Hence $d\log P_s>0$ for every $\sigma$, as in Result~\ref{prop:prices}.

Under demand-responsive provision, the price response in \eqref{eq:ces_price_response} determines the change in $s_E$ through \eqref{eq:dsE_eta}. Under fixed provision, $s_E$ also depends on output growth, as in \eqref{eq:dsE_fixed}. The other object in \eqref{eq:dTY} is the aggregate labor share, whose value-added weights are the same as in \eqref{eq:sL_identity}. Market production has weight $1-s_E$, public services have weight $s_E$, and the sectoral labor shares are now $\ell_m$ and $\ell_s$. Hence
\[
s_L=(1-s_E)\ell_m+s_E\ell_s.
\]
Therefore
\begin{equation}\label{eq:ces_dsL}
ds_L=(1-s_E)d\ell_m+s_Ed\ell_s+(\ell_s-\ell_m)ds_E,
\end{equation}
with
\begin{equation}\label{eq:ces_dell}
d\ell_j
=
\ell_j
\left[
\left(
-\frac{1}{1-I_j}
-\frac{(R/A_K)^{1-\sigma}-W^{1-\sigma}}{B_j}
\right)dI_j
+
(1-\sigma)(1-\ell_j)d\log W
\right].
\end{equation}
Define the CES counterparts of $D$ and $g$ in \eqref{eq:dsL_compact} as
\[
D_\sigma\equiv -\bigl[(1-s_E)d\ell_m+s_Ed\ell_s\bigr],
\qquad
g_\sigma\equiv \ell_s-\ell_m.
\]
Under CES task aggregation, the labor-intensity gap $s_L^s-s_L^m$ is $g_\sigma$. It generally differs from the threshold gap $g=I_m-I_s$, which equals the labor-intensity gap in the Cobb-Douglas case \eqref{eq:cost_shares} and enters the price response above.
With these definitions, \eqref{eq:ces_dsL} becomes
\[
ds_L=-D_\sigma+g_\sigma ds_E,
\]
and \eqref{eq:dTY} becomes
\begin{equation}\label{eq:ces_fiscal_decomp}
d(T/Y)=\bigl[1+\tau_c-(\tau_L-\tau_K)g_\sigma\bigr]ds_E+(\tau_L-\tau_K)D_\sigma.
\end{equation}
Equation~\eqref{eq:ces_fiscal_decomp} has the same structure as \eqref{eq:dTY_decomp}, except that sectoral labor shares $\ell_j$ replace the simple Cobb-Douglas form $1-I_j$. The CES analogue of $g$ is still positive. To see this, compare the two sectors at the same equilibrium values of $W$ and $R/A_K$. At those values, differences in $\ell_j$ across sectors come from differences in $I_j$. From \eqref{eq:ces_shares},
\[
\frac{\partial \ell_j}{\partial I_j}
=
-\frac{(R/A_K)^{1-\sigma}W^{1-\sigma}}{B_j^2}<0.
\]
Assumption~\ref{ass:initial} gives $I_s<I_m$, so $\ell_s>\ell_m$ and $g_\sigma>0$. Since $\ell_m,\ell_s\in(0,1)$, we have $g_\sigma\in(0,1)$. As in the Cobb-Douglas case, the aggregate labor share falls if and only if
\[
D_\sigma>g_\sigma ds_E.
\]

In the Cobb-Douglas limit, $\ell_j\to 1-I_j$ from \eqref{eq:cost_shares} and $d\ell_j\to -dI_j$. Hence
\[
D_\sigma\to (1-s_E)dI_m+s_EdI_s,
\qquad
g_\sigma\to I_m-I_s,
\]
which recovers $D$ and $g$ in \eqref{eq:dsL_compact}.

Since $B_m=1$, equation~\eqref{eq:ces_wage_norm} gives
\[
\ell_m=1-I_m(R/A_K)^{1-\sigma},
\qquad
d\ell_m=-(R/A_K)^{1-\sigma}\,dI_m.
\]
Market-sector automation therefore lowers the market-sector labor share for every $\sigma$. Combining this expression with the public-sector labor-share response in \eqref{eq:ces_dell} gives
\begin{multline}\label{eq:ces_D_explicit}
D_\sigma
=
(1-s_E)(R/A_K)^{1-\sigma}\,dI_m
+s_E\frac{(R/A_K)^{1-\sigma}W^{1-\sigma}}{B_s^2}\,dI_s \\
-(1-\sigma)s_E\ell_s(1-\ell_s)\,d\log W.
\end{multline}
The second term uses the expression for $\partial\ell_s/\partial I_s$ given above.

Under Assumption~\ref{ass:initial} and the automation changes considered in Section~\ref{sec:cs}, the first term is positive and the second is nonnegative. For $\sigma\geq1$, the last term is also nonnegative, so $D_\sigma>0$. For $\sigma<1$, higher wages raise the public-sector labor share at a given automation threshold. This effect can outweigh the reductions in labor shares caused by automation, so $D_\sigma$ need not be positive. At the calibrated values of Section~\ref{sec:cal}, with the initial thresholds, expenditure share and effective cost of capital $R/A_K$ held fixed, this cannot happen. With \eqref{eq:ces_wage_response} substituted for $d\log W$, the bounds $\ell_s(1-\ell_s)\leq1/4$, $0<1-(R/(A_KW))^{1-\sigma}<1$ and $(R/A_K)^{1-\sigma}>R/A_K$ for $\sigma<1$ show that \eqref{eq:ces_D_explicit} implies $D_\sigma>0$ for every $\sigma<1$ whenever $(1-s_E)R/A_K>s_E/[4(1-I_m)]$. This condition holds for the EU-27 and for every member state at the values in Tables~\ref{tab:eu_average} and~\ref{tab:cross_country_calibration}.

Because $g_\sigma\in(0,1)$, the net expenditure term in \eqref{eq:ces_fiscal_decomp} has the same sign as $ds_E$ for every $\sigma$. For $\sigma\geq1$, the displacement term $(\tau_L-\tau_K)D_\sigma$ is also positive, so Results~\ref{prop:fiscal_general} and~\ref{prop:demand} carry over with $g_\sigma$ and $D_\sigma$ in place of $g$ and $D$. For other parameter values with $\sigma<1$, $D_\sigma$ can be negative, and fiscal pressure need not rise even when $ds_E\geq0$.

\section{Sensitivity analyses}

\subsection{Initial cost saving, elasticity and tax measures}\label{app:saving_sensitivity}

Table~\ref{tab:scenario_grid_30} repeats the exercise in Table~\ref{tab:scenario_grid} with an initial cost saving of 30 percent instead of 20 percent, which raises $A_K$ for the EU-27 from 0.175 to 0.186. Every change in fiscal pressure keeps its sign. Under demand-responsive provision, the results differ from those in Table~\ref{tab:scenario_grid} by between 0.01 and 0.16 percentage points. The tax-base contribution, the larger of the two, hardly moves, because it is dominated by direct displacement, which does not depend on the cost saving. The expenditure contribution, in contrast, is about 1.6 times its value under the benchmark calibration. This is because the log responses of the wage and the public-service unit cost in \eqref{eq:dlogW} and \eqref{eq:dlogPs} are proportional to $\log(A_KW/R)$, whose initial value rises from $\log(1/0.8)\simeq0.22$ to $\log(1/0.7)\simeq0.36$. Under fixed provision, the larger saving raises output by more and therefore produces a larger fall in the expenditure share. The fall in fiscal pressure is larger by about 0.4 percentage points in the symmetric 10pp scenario.

Three further checks vary $\varepsilon$, the consumption-tax rate and the measure of capital-tax revenue. For $\varepsilon$ between 0.60 and 0.74, the EU demand-responsive effects differ from those under the benchmark calibration by less than 0.06pp. A consumption-tax rate of 0.142 instead of 0.196, at which a tax on the model's base, market-sector value added, raises the 10.5 percent of GDP collected in the EU \citep{ecTaxationTrends2026} once value added is converted to GDP at its observed ratio \citep{EurostatNama10a10}, changes them by about 0.01pp, because the tax enters only through the change in the expenditure share, which is small. Under fixed provision the expenditure share falls more, so the lower rate makes the fall in fiscal pressure up to 0.25pp smaller, but no EU result changes sign. Taxes on the stock of capital, such as property and wealth taxes, raised 2.2 of the 8.5 percent of GDP collected through capital taxes in the EU in 2023 \citep[Tables 61 and 53]{ecTaxationTrends2026}. Excluding these receipts from capital-tax revenue lowers $\tau_K$ to about 0.146. In the EU-27 scenarios, fiscal pressure then rises by up to 0.5pp more under demand-responsive provision and falls by up to 0.6pp less under fixed provision.

\begin{table}[htbp]
\centering
\caption{Change in fiscal pressure with a 30 percent initial cost saving, EU-27, 2023}
\label{tab:scenario_grid_30}
\footnotesize
\setlength{\tabcolsep}{3pt}\renewcommand{\arraystretch}{1.15}
\begin{tabular}{@{}l rr rrrr rrrr@{}}
\toprule
& & & \multicolumn{4}{c}{\textbf{Demand-responsive ($\boldsymbol{C_s}$)}} & \multicolumn{4}{c}{Fixed provision ($\widehat{C}_s$)} \\
\cmidrule(lr){4-7}\cmidrule(lr){8-11}
Scenario & $\Delta I_m$ & $\Delta I_s$ & Expenditure & \begin{tabular}[b]{@{}c@{}}Tax\\base\end{tabular} & $\boldsymbol{\Delta(T/Y)}$ & \begin{tabular}[b]{@{}c@{}}Output\\growth\\(\%)\end{tabular} & Expenditure & \begin{tabular}[b]{@{}c@{}}Tax\\base\end{tabular} & $\boldsymbol{\Delta(T/Y)}$ & \begin{tabular}[b]{@{}c@{}}Output\\growth\\(\%)\end{tabular} \\
\midrule
Private 1, Public 0 & 1 & 0 & 0.04 & 0.14 & 0.18 & 2.3 & $-$0.43 & 0.16 & $-$0.27 & 2.5 \\
Private 1, Public 0.5 & 1 & 0.5 & 0.03 & 0.16 & 0.18 & 2.5 & $-$0.52 & 0.18 & $-$0.34 & 2.8 \\
Symmetric 1 & 1 & 1 & 0.01 & 0.17 & 0.19 & 2.7 & $-$0.60 & 0.20 & $-$0.40 & 3.1 \\
Private 5, Public 0 & 5 & 0 & 0.20 & 0.70 & 0.90 & 12.5 & $-$2.20 & 0.83 & $-$1.37 & 14.3 \\
Private 5, Public 2.5 & 5 & 2.5 & 0.14 & 0.78 & 0.92 & 13.9 & $-$2.61 & 0.92 & $-$1.69 & 15.8 \\
Symmetric 5 & 5 & 5 & 0.08 & 0.87 & 0.95 & 15.2 & $-$3.01 & 1.01 & $-$2.01 & 17.3 \\
Private 10, Public 0 & 10 & 0 & 0.46 & 1.39 & 1.85 & 28.7 & $-$4.50 & 1.69 & $-$2.80 & 34.2 \\
Private 10, Public 5 & 10 & 5 & 0.32 & 1.56 & 1.88 & 32.2 & $-$5.28 & 1.86 & $-$3.42 & 37.9 \\
Symmetric 10 & 10 & 10 & 0.18 & 1.73 & 1.92 & 35.8 & $-$6.00 & 2.01 & $-$3.98 & 41.4 \\
\bottomrule
\end{tabular}
\begin{flushleft}
\scriptsize
Notes: As in Table~\ref{tab:scenario_grid}, with an initial cost saving of 30 percent ($A_K=0.186$).
\end{flushleft}
\end{table}

Table~\ref{tab:scenario_grid_publicled} reports the three public-led scenarios, in which public-sector automation increases by more than private-sector automation. Fiscal pressure rises under demand-responsive provision in all three, although the expenditure contribution is negative because public-service unit costs fall, and it falls under fixed provision. Result~\ref{prop:demand} does not cover these scenarios, so their signs depend on the calibrated parameters.

\begin{table}[htbp]
\centering
\caption{Change in fiscal pressure in the public-led scenarios, EU-27, 2023}
\label{tab:scenario_grid_publicled}
\footnotesize
\setlength{\tabcolsep}{3pt}\renewcommand{\arraystretch}{1.15}
\begin{tabular}{@{}l rr rrrr rrrr@{}}
\toprule
& & & \multicolumn{4}{c}{\textbf{Demand-responsive ($\boldsymbol{C_s}$)}} & \multicolumn{4}{c}{Fixed provision ($\widehat{C}_s$)} \\
\cmidrule(lr){4-7}\cmidrule(lr){8-11}
Scenario & $\Delta I_m$ & $\Delta I_s$ & Expenditure & \begin{tabular}[b]{@{}c@{}}Tax\\base\end{tabular} & $\boldsymbol{\Delta(T/Y)}$ & \begin{tabular}[b]{@{}c@{}}Output\\growth\\(\%)\end{tabular} & Expenditure & \begin{tabular}[b]{@{}c@{}}Tax\\base\end{tabular} & $\boldsymbol{\Delta(T/Y)}$ & \begin{tabular}[b]{@{}c@{}}Output\\growth\\(\%)\end{tabular} \\
\midrule
Private 1, Public 5 & 1 & 5 & $-$0.04 & 0.30 & 0.26 & 4.1 & $-$1.10 & 0.35 & $-$0.75 & 4.6 \\
Private 1, Public 10 & 1 & 10 & $-$0.11 & 0.46 & 0.35 & 6.3 & $-$1.74 & 0.52 & $-$1.23 & 7.0 \\
Private 5, Public 10 & 5 & 10 & $-$0.02 & 1.03 & 1.01 & 16.1 & $-$3.40 & 1.16 & $-$2.24 & 18.0 \\
\bottomrule
\end{tabular}
\begin{flushleft}
\scriptsize
Notes: As in Table~\ref{tab:scenario_grid}, with an initial cost saving of 20 percent ($A_K=0.175$).
\end{flushleft}
\end{table}

\subsection{Holding the capital stock fixed}\label{app:fixed_capital}

The baseline model allows firms to rent additional capital at the world rental rate $R$. Here we hold the capital stock fixed and apply the benchmark shock, which raises the automation threshold by 5pp in the private sector and 2.5pp in public services. Under Cobb-Douglas task aggregation, capital receives the share $I_j$ of the value of sector $j$'s output. The rental rate therefore rises until $RK=I_mY_m+I_sP_sY_s$, with $K$ held at its initial value. The wage still follows \eqref{eq:wage} and the expenditure share follows the provision rule, with $A_K$, tax rates and $\varepsilon$ unchanged. 

Table~\ref{tab:fixed_capital} compares the results, which Section~\ref{sec:capital_supply} discusses. The cost saving from replacing labor with capital, $1-R/(A_KW)$, falls from 20 percent to about 4 percent under demand-responsive provision and 3 percent under fixed provision. The expenditure share falls under fixed provision, because service costs fall by a larger percentage than value added measured in units of the private good.

The same calculation works for every scenario in Table~\ref{tab:scenario_grid} with a private-sector increase of 1pp or 5pp. The three 10pp scenarios have no solution with $R/A_K<1$. At $R/A_K=1$, labor and capital cost the same in automatable tasks. Some of those tasks can then remain with labor, which requires a separate calculation that we do not report here.

\begin{table}[htbp]
\centering
\caption{Automation with a fixed rental rate or a fixed capital stock, EU-27, 2023}
\label{tab:fixed_capital}
\footnotesize
\begin{tabular}{@{}l rr rr@{}}
\toprule
& \multicolumn{2}{c}{\textbf{Demand-responsive ($\boldsymbol{C_s}$)}} & \multicolumn{2}{c}{Fixed provision ($\widehat{C}_s$)} \\
\cmidrule(lr){2-3}\cmidrule(lr){4-5}
& Fixed $R$ & Fixed $K$ & Fixed $R$ & Fixed $K$ \\
\midrule
Rental rate $R$ & 0.158 & 0.172 & 0.158 & 0.173 \\
Capital stock (\%) & 23 & 0 & 27 & 0 \\
Wage (\%) & 2.8 & $-$9.4 & 2.8 & $-$9.9 \\
Unit cost of public services (\%) & 1.5 & $-$5.6 & 1.5 & $-$5.9 \\
Real value added (\%) & 12.3 & 0.6 & 14.0 & 0.6 \\
Value added in private-good units (\%) & 12.6 & $-$0.5 & 14.3 & $-$0.5 \\
Expenditure contribution & 0.09 & $-$0.34 & $-$2.45 & $-$1.19 \\
Tax-base contribution & 0.79 & 0.81 & 0.91 & 0.85 \\
$\Delta(T/Y)$ & 0.87 & 0.47 & $-$1.54 & $-$0.34 \\
\bottomrule
\end{tabular}
\begin{flushleft}
\scriptsize
Notes: The automation threshold rises by 5pp in the private sector and 2.5pp in public services, the benchmark scenario. Parameters are from Table~\ref{tab:eu_average}, with $A_K$ set for a 20 percent initial cost saving. The fixed-$R$ columns keep the rental rate at 0.158. Their fiscal results and real output growth match Table~\ref{tab:scenario_grid}, and their wage and capital-stock changes match Section~\ref{sec:quant}, where capital income and the capital stock grow at the same rate because $R$ is fixed. In the fixed-$K$ columns, the stock stays at its initial level and the rental rate adjusts until firms' demand equals that stock. The first row shows the rental rate after automation. Rows marked (\%) show percentage changes, with real value added measured by a Fisher quantity index. The last three rows show changes in percentage points of value added, as defined in Table~\ref{tab:scenario_grid}.
\end{flushleft}
\end{table}

\section{Country-level calibration and results}\label{app:countries}

Table~\ref{tab:cross_country_calibration} reports the calibrated parameters for each member state, and Table~\ref{tab:cross_country_results} reports the country results discussed in Section~\ref{sec:cross_country}.

\paragraph{Sensitivity of the country ranking.}

When taxes on the stock of capital are excluded from capital-tax revenue, as in Appendix~\ref{app:saving_sensitivity}, the country ranking by the demand-responsive increase in the benchmark scenario has a rank correlation of 0.90 relative to the ranking under the benchmark calibration. Under the same shock, the changes caused by the exclusion are larger in some countries than for the EU-27, reaching about 0.5pp in France under fixed provision. No reported fiscal-pressure result changes sign. A second variant uses the country estimates of $\varepsilon$ in \citet{francois2023intratemporal} where available, which gives a rank correlation of 0.93 with the ranking under the benchmark calibration.\footnote{The estimates cover 14 of the 27 member states. We keep 0.67 for the remaining member states and for Portugal, whose negative estimate lies outside the range a CES aggregator allows. We use all other estimates as reported, including Belgium's, which is close to zero. The fixed-provision results do not depend on $\varepsilon$.} Fiscal pressure still rises in every demand-responsive scenario in Table~\ref{tab:cross_country_results}, but Ireland falls from fifth to fifteenth and Denmark from eighteenth to twenty-third, because their estimates exceed one.

The classification of payroll taxes and mixed income also matters. Payroll taxes were 0.7 percent of GDP in the EU in 2023 but 9.1 percent in Sweden, where most employer charges on wages are recorded as payroll taxes rather than as part of compensation of employees \citep{EurostatGov10aTaxag,SCB2025GNI}. We move them from capital income to labor income in each sector in proportion to compensation. Because capital-tax revenue is unchanged, this also raises $\tau_K$ and narrows the labor--capital tax differential. The demand-responsive increase in the benchmark scenario then falls from 1.10pp to 0.91pp in Sweden, almost entirely because of the higher $\tau_K$, and from 0.87pp to 0.86pp in the EU-27 (Table~\ref{tab:cross_country_results}). The country ranking as a whole hardly moves, with a rank correlation of 0.98 relative to the ranking under the benchmark calibration.

The treatment of mixed income affects both the capital-tax base and measured labor shares. In a variant, we treat two thirds of mixed income as private-sector labor income and move the same fraction of the taxes on self-employment income from capital to labor (for Bulgaria, which lacks 2023 data, we use its 2022 ratio of mixed income to GDP).\footnote{Mixed income is the gross mixed income (B.3g) of households and non-profit institutions serving households in Eurostat table nasa\_10\_nf\_tr \citeyearpar{EurostatNasa10NfTr}, extracted on 1 November 2026. Taxes on self-employment income are from Table 59 of the Taxation Trends data \citep{ecTaxationTrends2026}.} For the same shock, the EU-27 demand-responsive increase then falls from 0.87pp to 0.83pp. No result in Table~\ref{tab:cross_country_results} changes sign under either variant, but the mixed-income variant shifts the country ranking considerably, with a rank correlation of 0.76 relative to the ranking under the benchmark calibration. Greece, where mixed income is 22 percent of GDP and the taxes recorded on it less than 1 percent of GDP, falls from second to twenty-third. Moving this lightly taxed income to labor lowers Greece's $\tau_L$ and raises its $\tau_K$, which more than halves its labor--capital tax differential.

\begin{table}[htbp]
\centering
\caption{Calibrated parameters by country, 2023}
\label{tab:cross_country_calibration}
\begin{tabular}{l r rr rrr r}
\toprule
Country & $s_E$ & $I_m$ & $I_s$ & $\tau_L$ & $\tau_K$ & $\tau_c$ & $A_K$ \\
\midrule
Austria & 0.181 & 0.499 & 0.252 & 0.404 & 0.198 & 0.216 & 0.177 \\
Belgium & 0.216 & 0.558 & 0.133 & 0.403 & 0.248 & 0.187 & 0.174 \\
Bulgaria & 0.162 & 0.554 & 0.225 & 0.248 & 0.145 & 0.235 & 0.175 \\
Croatia & 0.161 & 0.520 & 0.200 & 0.284 & 0.146 & 0.280 & 0.176 \\
Cyprus & 0.184 & 0.559 & 0.227 & 0.338 & 0.201 & 0.203 & 0.174 \\
Czechia & 0.156 & 0.582 & 0.266 & 0.355 & 0.133 & 0.220 & 0.173 \\
Denmark & 0.195 & 0.495 & 0.116 & 0.360 & 0.218 & 0.281 & 0.177 \\
Estonia & 0.167 & 0.488 & 0.190 & 0.339 & 0.074 & 0.267 & 0.177 \\
Finland & 0.211 & 0.509 & 0.211 & 0.395 & 0.194 & 0.249 & 0.176 \\
France & 0.213 & 0.478 & 0.246 & 0.390 & 0.274 & 0.198 & 0.178 \\
Germany & 0.191 & 0.471 & 0.207 & 0.360 & 0.185 & 0.176 & 0.178 \\
Greece & 0.175 & 0.668 & 0.275 & 0.433 & 0.156 & 0.216 & 0.170 \\
Hungary & 0.158 & 0.565 & 0.285 & 0.349 & 0.151 & 0.287 & 0.174 \\
Ireland & 0.096 & 0.770 & 0.187 & 0.324 & 0.100 & 0.209 & 0.166 \\
Italy & 0.155 & 0.622 & 0.302 & 0.442 & 0.199 & 0.191 & 0.172 \\
Latvia & 0.187 & 0.480 & 0.238 & 0.319 & 0.080 & 0.251 & 0.177 \\
Lithuania & 0.154 & 0.515 & 0.194 & 0.310 & 0.114 & 0.211 & 0.176 \\
Luxembourg & 0.181 & 0.497 & 0.230 & 0.345 & 0.272 & 0.281 & 0.177 \\
Malta & 0.144 & 0.621 & 0.194 & 0.239 & 0.127 & 0.171 & 0.172 \\
Netherlands & 0.209 & 0.548 & 0.234 & 0.302 & 0.231 & 0.226 & 0.175 \\
Poland & 0.149 & 0.624 & 0.244 & 0.323 & 0.192 & 0.210 & 0.172 \\
Portugal & 0.181 & 0.518 & 0.203 & 0.304 & 0.187 & 0.190 & 0.176 \\
Romania & 0.131 & 0.632 & 0.192 & 0.297 & 0.092 & 0.180 & 0.172 \\
Slovakia & 0.161 & 0.605 & 0.218 & 0.383 & 0.092 & 0.218 & 0.173 \\
Slovenia & 0.165 & 0.471 & 0.204 & 0.349 & 0.138 & 0.250 & 0.178 \\
Spain & 0.172 & 0.549 & 0.146 & 0.361 & 0.197 & 0.156 & 0.175 \\
Sweden & 0.208 & 0.519 & 0.258 & 0.375 & 0.149 & 0.260 & 0.176 \\
\bottomrule
\end{tabular}
\begin{flushleft}
\scriptsize
Sources and definitions as in Table~\ref{tab:eu_average}. Value added and compensation for 2023 are provisional for Germany, Greece and Spain.
\end{flushleft}
\end{table}

\begin{table}[htbp]
\centering
\caption{Change in fiscal pressure by country in the benchmark scenario, with and without its public-sector part, 2023}
\label{tab:cross_country_results}
\small
\begin{tabular}{l r rrr r rr}
\toprule
& & \multicolumn{4}{c}{Demand-responsive ($C_s$)} & \multicolumn{2}{c}{Fixed provision ($\widehat{C}_s$)} \\
\cmidrule(lr){3-6}\cmidrule(lr){7-8}
& & \multicolumn{3}{c}{$\Delta I_s=2.5$} & $\Delta I_s=0$ & $\Delta I_s=2.5$ & $\Delta I_s=0$ \\
\cmidrule(lr){3-5}
Country & $\tau_L-\tau_K$ & Expenditure & Tax base & $\Delta(T/Y)$ & $\Delta(T/Y)$ & $\Delta(T/Y)$ & $\Delta(T/Y)$ \\
\midrule
Slovakia & 0.291 & 0.10 & 1.33 & 1.43 & 1.34 & $-$0.98 & $-$0.81 \\
Greece & 0.277 & 0.13 & 1.25 & 1.38 & 1.29 & $-$1.77 & $-$1.52 \\
Estonia & 0.265 & 0.08 & 1.21 & 1.29 & 1.21 & $-$0.78 & $-$0.61 \\
Italy & 0.243 & 0.09 & 1.11 & 1.20 & 1.14 & $-$1.28 & $-$1.10 \\
Ireland & 0.224 & 0.15 & 1.05 & 1.20 & 1.17 & $-$1.28 & $-$1.16 \\
Latvia & 0.239 & 0.07 & 1.08 & 1.15 & 1.08 & $-$1.13 & $-$0.91 \\
Czechia & 0.222 & 0.08 & 1.02 & 1.10 & 1.05 & $-$1.22 & $-$1.04 \\
Sweden & 0.226 & 0.09 & 1.01 & 1.10 & 1.02 & $-$1.65 & $-$1.36 \\
Romania & 0.205 & 0.10 & 0.95 & 1.05 & 1.01 & $-$1.05 & $-$0.91 \\
Slovenia & 0.211 & 0.07 & 0.96 & 1.03 & 0.98 & $-$0.96 & $-$0.78 \\
Austria & 0.206 & 0.07 & 0.93 & 1.01 & 0.95 & $-$1.24 & $-$1.03 \\
Hungary & 0.198 & 0.08 & 0.91 & 0.99 & 0.94 & $-$1.44 & $-$1.23 \\
Finland & 0.201 & 0.09 & 0.89 & 0.99 & 0.92 & $-$1.70 & $-$1.39 \\
Lithuania & 0.196 & 0.08 & 0.90 & 0.98 & 0.93 & $-$0.98 & $-$0.81 \\
Germany & 0.175 & 0.07 & 0.79 & 0.86 & 0.81 & $-$1.30 & $-$1.07 \\
Spain & 0.164 & 0.10 & 0.74 & 0.84 & 0.80 & $-$1.36 & $-$1.14 \\
Belgium & 0.155 & 0.12 & 0.68 & 0.81 & 0.76 & $-$2.04 & $-$1.68 \\
Denmark & 0.142 & 0.10 & 0.64 & 0.74 & 0.71 & $-$1.74 & $-$1.43 \\
Croatia & 0.138 & 0.08 & 0.63 & 0.71 & 0.69 & $-$1.51 & $-$1.28 \\
Cyprus & 0.137 & 0.09 & 0.62 & 0.71 & 0.69 & $-$1.86 & $-$1.59 \\
Poland & 0.131 & 0.10 & 0.60 & 0.70 & 0.68 & $-$1.79 & $-$1.58 \\
Malta & 0.112 & 0.10 & 0.52 & 0.62 & 0.60 & $-$1.69 & $-$1.48 \\
Portugal & 0.117 & 0.08 & 0.53 & 0.61 & 0.59 & $-$1.69 & $-$1.43 \\
France & 0.116 & 0.08 & 0.52 & 0.59 & 0.57 & $-$1.93 & $-$1.60 \\
Bulgaria & 0.103 & 0.09 & 0.47 & 0.56 & 0.55 & $-$1.79 & $-$1.55 \\
Netherlands & 0.071 & 0.10 & 0.32 & 0.42 & 0.42 & $-$2.53 & $-$2.14 \\
Luxembourg & 0.073 & 0.08 & 0.33 & 0.41 & 0.42 & $-$2.02 & $-$1.72 \\
\midrule
EU-27 & 0.174 & 0.09 & 0.79 & 0.87 & 0.83 & $-$1.54 & $-$1.29 \\
\bottomrule
\end{tabular}
\begin{flushleft}
\scriptsize
Notes: The column $\tau_L-\tau_K$ gives the labor--capital tax differential as a difference in tax rates, which the text and Figure~\ref{fig:cross_country_bar} express in percentage points. Expenditure is $(1+\tau_c)\Delta s_E$ and tax base is $-(\tau_L-\tau_K)\Delta s_L$, as in Table~\ref{tab:scenario_grid}. All other entries are exact finite changes in percentage points of value added for $\Delta I_m=5$, with public automation of 2.5pp (the benchmark scenario) or zero, using each country's own initial shares and tax rates, a common $R$ and $\varepsilon$, and $A_K$ set for a 20 percent initial cost saving. Components may not sum to the total because of rounding. The EU-27 row uses the parameters in Table~\ref{tab:eu_average} and the country rows those in Table~\ref{tab:cross_country_calibration}. Countries are sorted by the demand-responsive response in the benchmark scenario. Ireland's low private-sector labor share partly reflects multinational profit shifting \citep{torslov2023missing}.
\end{flushleft}
\end{table}

\end{document}